\documentclass[journal]{IEEEtran}
\usepackage{cite}
\usepackage{graphicx}
\usepackage{float}
\usepackage{subfloat}
\usepackage{stfloats}
\ifCLASSOPTIONcompsoc
\usepackage[caption=false,font=normalsize,labelfont=sf,textfont=sf]{subfig}
\else
\usepackage[caption=false,font=footnotesize]{subfig}
\fi
\usepackage{amsmath}
\usepackage{amsthm}
\usepackage{algorithmic}
\usepackage{array}
\usepackage{tikz}
\usepackage{color}
\usepackage{epsfig,latexsym}
\usepackage{indentfirst}
\usepackage{amssymb}
\usepackage{times}
\usepackage{mathtools}
\usepackage{psfrag}
\usepackage{lastpage}
\usepackage{fancyhdr}
\usepackage{color}
\usepackage{amsthm}
\usepackage{bigints}
\usepackage{algorithm}
\usepackage{algorithmic}
\usepackage{setspace}
\usepackage{booktabs}
\usepackage{amsfonts}
\usepackage{epstopdf}
\usepackage{cleveref}
\usepackage{diagbox}
\usepackage{multirow}
\usepackage{flushend}

\theoremstyle{remark}
\newtheorem{theorem}{\quad Theorem}

\allowdisplaybreaks[4]

\begin{document}

\title{Reconfigurable Intelligent Surfaces Relying on Non-Diagonal Phase Shift Matrices}
\author{
\thanks{L. Hanzo would like to acknowledge the financial support of the
Engineering and Physical Sciences Research Council projects EP/P034284/1
and EP/P003990/1 (COALESCE) as well as of the European Research
Council's Advanced Fellow Grant QuantCom (Grant No. 789028). The support of Interdigital is also gratefully acknowledged. \textit{(Corresponding author: Lajos Hanzo.)}}

Qingchao Li,  Mohammed El-Hajjar, \textit{Senior Member, IEEE},  Ibrahim Hemadeh, \textit{Member, IEEE},  Arman Shojaeifard, \textit{Senior Member, IEEE},  Alain A. M. Mourad,  Bruno Clerckx, \textit{Senior Member, IEEE},  Lajos Hanzo, \textit{Fellow, IEEE}

\thanks{Qingchao Li, Mohammed El-Hajjar and Lajos Hanzo are with the Electronics and Computer Science, University of Southampton, Southampton SO17 1BJ, U.K. (e-mail: Qingchao.Li@soton.ac.uk; meh@ecs.soton.ac.uk; lh@ecs.soton.ac.uk).

Ibrahim Hemadeh, Arman Shojaeifard and Alain A. M. Mourad are with InterDigital, London EC2A 3QR, U.K. (e-mail: Ibrahim.Hemadeh@InterDigital.com; arman.shojaeifard@interdigital.com; Alain.Mourad@interdigital.com).

Bruno Clerckx is with the Communications and Signal Processing Group, Department of Electrical and Electronic Engineering, Imperial College London, London SW7 2AZ, U.K. (email: b.clerckx@imperial.ac.uk).
}}

\maketitle

\begin{abstract}
Reconfigurable intelligent surfaces (RIS) have been actively researched as a potential technique for future wireless communications, which intelligently ameliorate the signal propagation environment. In the conventional design, each RIS element configures and reflects its received signal independently of all other RIS elements, which results in a diagonal phase shift matrix. By contrast, we propose a novel RIS architecture, where the incident signal impinging on one element can be reflected from another element after an appropriate phase shift adjustment, which increases the flexibility in the design of RIS phase shifts, hence, potentially improving the system performance. The resultant RIS phase shift matrix also has off-diagonal elements, as opposed to the pure diagonal structure of the conventional design. Compared to the state-of-art fully-connected/group-connected RIS structures, our proposed RIS architecture has lower complexity, while attaining a higher channel gain than the group-connected RIS structure, and approaching that of the fully-connected RIS structure. We formulate and solve the problem of maximizing the achievable rate of our proposed RIS architecture by jointly optimizing the transmit beamforming and the non-diagonal phase shift matrix based on alternating optimization and semi-define relaxation (SDR) methods. Moreover, the closed-form expressions of the channel gain, the outage probability and bit error ratio (BER) are derived. Simulation results demonstrate that our proposed RIS architecture results in an improved performance in terms of the achievable rate compared to the conventional architecture, both in single-user as well as in multi-user scenarios.
\end{abstract}
\begin{IEEEkeywords}
Reconfigurable intelligent surfaces (RIS), channel gain, outage probability, average bit error ratio (BER), joint beamforming.
\end{IEEEkeywords}

\section{Introduction}
\IEEEPARstart{I}{n} future wireless networks an ultra-high data rate, ultra-low latency, ultra-high reliability and ubiquitous connectivity is required for communication, computation, sensing and location awareness, especially in the Internet of Things (IoT) \cite{ma2019high,yetgin2017survey}. Hence Boccardi \textit{et al.} \cite{boccardi2014five} identified a range of sophisticated enabling techniques, including massive multiple-input-multiple-output (MIMO) solutions and millimeter wave communications. As an additional promising component, reconfigurable intelligent surfaces (RIS) have also been proposed for future wireless systems to intelligently reconfigure the propagation environment \cite{elmossallamy2020reconfigurable,di2020smart,gong2020toward,pan2020multicell,dai2020reconfigurable,hou2020reconfigurable,zhou2020joint,cao2021ai}. Explicitly, in RIS, a large number of passive scattering elements are employed for creating additional signal propagation paths between the base station (BS) and the mobile terminal users, which can substantially enhance the performance, especially when the direct link between the BS and the users is blocked.

Previous contributions on RIS are mainly focused on maximizing the spectral efficiency/achievable rate or minimizing the transmission power \cite{wu2019intelligent,ning2020beamforming,wang2020intelligent,han2019large,wang2021joint,wu2019beamforming,zhang2020beyond,chen2019sum,xu2019discrete,lin2020reconfigurable}. In \cite{wu2019intelligent}, Wu and Zhang minimized the transmission power in the downlink of RIS-aided multi-user MIMO systems, where the popular alternating optimization and semi-define relaxation (SDR) methods were employed for jointly optimizing the active transmission beamforming (TBF) of the BS and the passive beamforming, represented by the RIS phase shift matrix. This was achieved by approximately configuring the RIS reflecting elements. Ning \textit{et al.} \cite{ning2020beamforming} maximized the sum-path-gain of RIS-assisted point-to-point MIMO systems, where the low-complexity alternating direction method of multipliers (ADMM) was employed for configuring the RIS phase shift matrix, while the classic singular value decomposition (SVD) was employed for designing the TBF. In \cite{wang2020intelligent}, the optimal closed-form solution of the phase shift matrix and TBF were derived by Wang \textit{et al.} for single-user multiple-input-single-output (MISO) millimeter wave systems.

\begin{table*}[!b]
\setstretch{1}
\vspace{-0em}
\footnotesize
\setlength{\abovecaptionskip}{1em}
\setlength{\belowcaptionskip}{0em}
\begin{center}
\caption{Novelty comparison with the literature.}\label{Table_1}
\begin{tabular}{*{16}{l}}
\toprule
     & Our paper & \cite{wu2019intelligent} & \cite{ning2020beamforming} & \cite{wang2020intelligent} & \cite{han2019large} & \cite{wang2021joint} & \cite{wu2019beamforming} & \cite{zhang2020beyond} & \cite{chen2019sum} & \cite{xu2019discrete} & \cite{lin2020reconfigurable} & \cite{basar2019wireless} & \cite{yang2020accurate} & \cite{shen2020modeling}\\
\midrule
\midrule
    Beamforming design & $\surd$ & $\surd$ & $\surd$ & $\surd$ & $\surd$ & $\surd$ & $\surd$ & $\surd$ & $\surd$ & $\surd$ & $\surd$ &  &  &  $\surd$\\
\midrule
    Outage performance analysis & $\surd$ &  &  &  &  &  &  &  &  &  &  &  & $\surd$  & \\
\midrule
    Average BER analysis & $\surd$ &  &  &  &  &  &  &  &  &  &  & $\surd$ & $\surd$  & \\
\midrule
    Multi-user & $\surd$ & $\surd$ &  &  &  &  & $\surd$ &  & $\surd$ &  &  &  &   & \\
\midrule
    Cooperation among RIS elements & $\surd$ &  &  &  &  &  &  &  &  &  &  &  &   & $\surd$\\
\bottomrule
\end{tabular}
\end{center}
\vspace{-0em}
\end{table*}

Additionally, in order to reduce the overhead of channel estimation, Han \textit{et al.} \cite{han2019large} maximized the ergodic spectral efficiency of RIS-assisted systems communicating over Rician fading channels, relying on the angle of arrival (AoA) and angle of departure (AoD) information. The problem of maximizing the ergodic spectral efficiency based on statistical CSI in Rician fading channels was studied in \cite{wang2021joint}, where Wang \textit{et al.} considered the effect of channel correlation on the ergodic spectral efficiency.

While considering quantized RIS phase shifts, Wu and Zhang \cite{wu2019beamforming} employed the popular branch-and-bound method and an exhaustive search method for single-user and multi-user RIS-assisted systems, respectively. The branch-and-bound algorithm was also employed by Zhang \textit{et al.} \cite{zhang2020beyond} to design a discrete phase shift matrix, where the RIS has the dual functions of both reflection and refraction. In \cite{chen2019sum}, the local search (LS) method and cross-entropy (CE) method were proposed for optimizing the RIS phase shift matrices having discrete entries. In \cite{xu2019discrete}, Xu \textit{et al.} designed their discrete phase shift matrix based on low resolution digital-to-analog converters, and derived the lower bound of the asymptotic rate. Furthermore, the problem of maximizing the achievable rate of RIS users was studied by Lin \textit{et al.} \cite{lin2020reconfigurable}, where the novel concept of reflection pattern modulation was employed.

The theoretical performance analysis of RIS-assisted single-input-single-output (SISO) systems was also investigated. In \cite{basar2019wireless}, the theoretical channel gain of RIS-aided systems was characterized by Basar \textit{et al.}, compared to that of conventional SISO systems operating without RIS. Furthermore, the instantaneous signal-noise-ratio (SNR) has been derived based on the central-limit-theorem (CLT). In \cite{yang2020accurate}, Yang \textit{et al.} derived the accurate closed-form theoretical instantaneous SNR expression for a dual-hop RIS-aided scheme.

However, the RIS structures of \cite{wu2019intelligent,ning2020beamforming,wang2020intelligent,han2019large,wang2021joint,wu2019beamforming,zhang2020beyond,chen2019sum,xu2019discrete,lin2020reconfigurable,basar2019wireless,yang2020accurate}, assumed that the incident signal impinging on a specific element can be only reflected from the same element after phase shift adjustment. In other words, there was no controlled relationship among the RIS elements. We refer to this RIS architecture as the conventional RIS architecture. Therefore, the phase shift matrix in these designs has a diagonal structure, which does not exploit the full potential of RIS for enhancing the system performance. To the best of our knowledge, only Shen \textit{et al.} \cite{shen2020modeling} studied the cooperation among RIS elements, where fully-connected/group-connected network architectures were proposed. The associated theoretical analysis and simulation results demonstrated that their architectures are capable of significantly increasing the received signal power, compared to the conventional RIS structure, when considering SISO systems. However, the performance enhancement reported in \cite{shen2020modeling} is attained at the cost of increased optimization complexity. For example, $\frac{N(N+1)}{2}$ entries are available in the fully-connected phase shift matrix and $\frac{N(G+1)}{2}$ entries are in the group-connected phase shift matrix. By contrast, there are only $N$ non-zero entries in the conventional RIS case, where $N$ is the number of RIS elements and $G$ is the group size of the group-connected architecture. This increases the number of entries to optimize in the phase shift matrix and the amount of information to be transferred over the BS-RIS control link. Furthermore, the fully-connected/group-connected phase shift matrix of \cite{shen2020modeling} has the additional constraint of symmetry because all the proposed architectures are reciprocal.

By contrast, we propose a novel RIS structure relying on non-reciprocal connections, in which the signal impinging on a specific element can be reflected from another element after phase shift adjustment, so the phase shift matrix can be non-symmetric and of non-diagonal nature. These provide flexibility in terms of configuring the RIS structure for enhancing the system performance. We employ alternating optimization and SDR methods for jointly optimizing the TBF and phase shift matrix for single-user MISO systems and multi-user MIMO systems, respectively. The theoretical analysis and simulation results demonstrate that our proposed RIS architecture achieves better channel gain, outage probability, average bit error ratio (BER) and throughput than the conventional RIS architecture. Furthermore, in our proposed RIS architecture, there are only $N$ non-zero entries in the phase shift matrix, which is the same as that in the conventional RIS architecture. Additionally, the position of the non-zero entries in our proposed RIS architecture has to be updated, which requires $N$ values of information. Hence the total information to be exchanged over the BS-RIS control link in each coherence time duration of our proposed RIS architecture is $2N$, i.e. significantly lower than that of the fully-connected RIS architecture. Against this background, the novel contributions of this paper are summarized as follows:
\begin{itemize}
  \item We propose a novel RIS architecture having a non-diagonal phase shift matrix, and jointly design the TBF and phase shift matrix by alternating optimization and SDR methods for maximizing the achievable rate.
  \item We provide both theoretical and simulation results for characterizing the performance of our proposed RIS architecture, which is better than the conventional RIS architecture in terms of its channel gain, outage probability, average BER and achievable rate.
  \item We show that the performance of our proposed architecture approaches that of the state-of-the-art fully-connected RIS architecture, while providing better performance than that of the group-connected RIS architecture, when the number of RIS elements increases. Additionally, this is attained while requiring a reduced information exchange over the BS-RIS control link and fewer optimized phase shift entries than the fully-connected architecture.
\end{itemize}

Finally, Table \ref{Table_1} explicitly contrasts our contributions to the literature.

The rest of this paper is organized as follows. In Section \ref{Section_System_Model}, we present the system model. The beamforming design methods are formulated in Section \ref{Section_Beamforming_Design}. Our theoretical analysis and simulation results are presented in Section \ref{Section_Theoretical_Analysis} and Section \ref{Section_Performance_results_and_analysis}, respectively. Finally, we conclude in Section \ref{Section_Conclusions}.

\textit{Notations:} Vectors and matrices are denoted by boldface lower and upper case letters, respectively. $(\cdot)^{\mathrm{T}}$, $(\cdot)^{\mathrm{H}}$ represent the operation of transpose and Hermitian transpose, respectively. $\mathbb{C}^{m\times n}$ denotes the space of $m\times n$ complex-valued matrix. $[\mathbf{a}]_{i}$ represents the $i$th element in vector $\mathbf{a}$, and $[\mathbf{A}]_{i,j}$ represents the $(i,j)$th element in matrix $\mathbf{A}$. $\mathrm{diag}\{\mathbf{a}\}$ denotes a diagonal matrix with each element being the elements in vector $\mathbf{a}$, $\mathbf{I}$ represents the identity matrix. $\mathrm{Tr(\mathbf{A})}$, $\mathrm{Rank}(\mathbf{A})$ and $|\mathbf{A}|$ represent the trace, rank and determinant of matrix $\mathbf{A}$, respectively. $\mathbf{A}\succeq0$ indicates that $\mathbf{A}$ is a positive semi-define matrix. $|a|$ (or $|\mathbf{a}|$) and $\angle a$ (or $\angle\mathbf{a}$) represent the amplitude and phase of the complex scalar $a$ (or complex vector $\mathbf{a}$), respectively. $\|\mathbf{a}\|$ denotes the 2-norm of vector $\mathbf{a}$. $f_X(x)$ and $F_X(x)$ are the probability density function (PDF) and cumulative distribution function (CDF) of random variables $X$. A circularly symmetric complex Gaussian random vector with mean $\mathbf{\mu}$ and covariance matrix $\mathbf{\Sigma}$ is denoted as $\mathcal{CN}(\mathbf{\mu},\mathbf{\Sigma})$. $\mathbb{E}(X)$ represents the mean of the random variable $X$.

\section{System Model}\label{Section_System_Model}
\vspace{-2mm}
\subsection{Channel Model}
\begin{figure}[!t]
\vspace{-0mm}
\setstretch{1}
\setlength{\abovecaptionskip}{-3pt}
\setlength{\belowcaptionskip}{-3pt}
    \centering
    \includegraphics[width=2in]{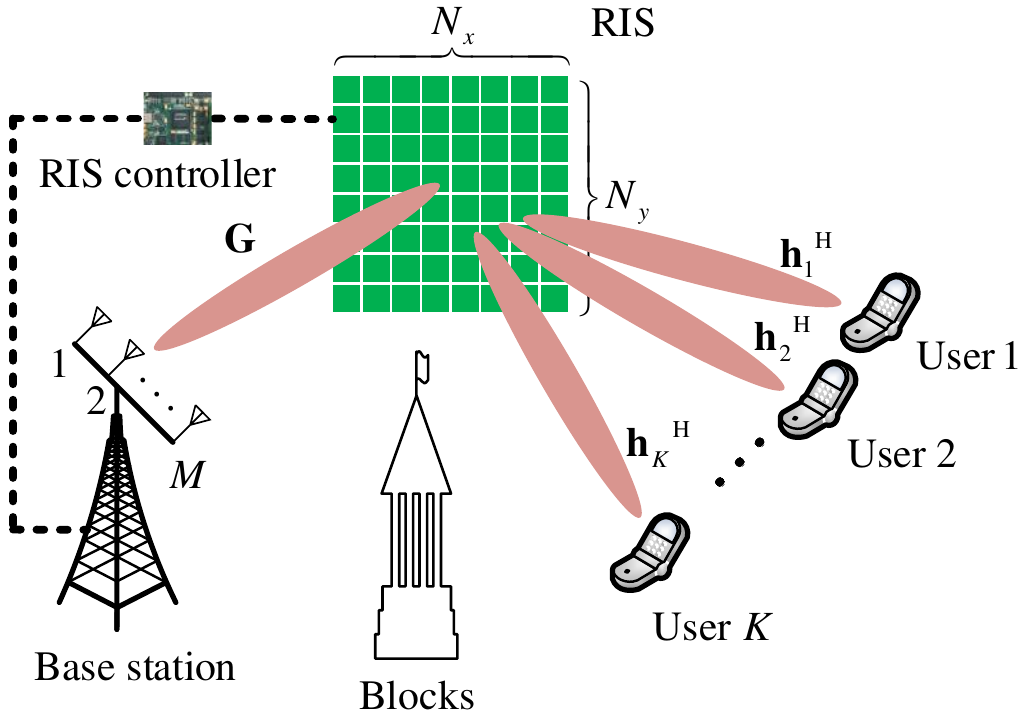}
    \caption{The employed RIS-assisted wireless communication system model, including a $M$-antenna base station, $K$ single-antenna users and a RIS with $N=N_x\times N_y$ elements.}\label{Figure_1}
\vspace{-5mm}
\end{figure}

The RIS-assisted system model is illustrated in Fig. \ref{Figure_1}, including a BS having $M$ transmit antennas, $K$ single-antenna mobile receivers, and a RIS with $N$ elements. The direct link between the BS and the users is blocked, while the RIS creates additional communication links arriving from the BS to the users. The link spanning from the BS to the RIS is denoted as $\mathbf{G}=[\mathbf{g}_1,\mathbf{g}_2,\cdots,\mathbf{g}_M]\in\mathbb{C}^{N\times M}$, where $\mathbf{g}_m\in\mathbb{C}^{N\times 1}$ ($m=1,2,\cdots,M$) represents the channel vector from the $m$th BS antenna to the RIS. The links impinging from the RIS to the users are denoted as $\mathbf{H}=[\mathbf{h}_{1},\mathbf{h}_{2},\cdots,\mathbf{h}_{K}]^\mathrm{H}\in\mathbb{C}^{K\times N}$, where $\mathbf{h}_{k}^{\mathrm{H}}\in\mathbb{C}^{1\times N}$ ($k\in{1,2,\cdots,K}$) represents the channel vector from the RIS to the $k$th single-antenna receiver. We employ the far field RIS channel model, since the size of the RIS is negligible compared to both the BS-RIS distance and to the RIS-user distance \cite{wu2021intelligent}. Additionally, we consider a distance-dependent path loss model, and Rician channel model for the small scaling fading \cite{wu2019intelligent}.

The Rician channel model from the BS to the RIS is given be
\begin{align}\label{Rician_model_1}
    \mathbf{G}=\sqrt{\frac{\kappa_{\mathbf{G}}}{1+\kappa_{\mathbf{G}}}}\overline{\mathbf{G}}+\sqrt{\frac{1}{1+\kappa_{\mathbf{G}}}}\widetilde{\mathbf{G}},
\end{align}
where $\kappa_{\mathbf{G}}$ is the Rician factor, $\overline{\mathbf{G}}\in\mathbb{C}^{N\times M}$ and $\widetilde{\mathbf{G}}\in\mathbb{C}^{N\times M}$ represent the line-of-sight (LoS) and non-line-of-sight (NLoS) components, respectively.

The LoS component $\overline{\mathbf{G}}$ is expressed as
\begin{align}\label{Rician_model_2}
    \overline{\mathbf{G}}=\sqrt{\varrho_t}\mathbf{f}_{\mathrm{RIS}}^\mathrm{A}\left(\phi^{\mathrm{A}},\varphi^{\mathrm{A}}\right){\mathbf{f}_{\mathrm{BS}}^\mathrm{D}}^{\mathrm{H}}\left(\psi^{\mathrm{D}}\right),
\end{align}
where $\varrho_t=C_0d_t^{-\alpha_t}$ denotes the path loss of the BS-RIS link, in which $d_t$ denotes the distance between the BS and the RIS. $C_0$ is the path loss at the reference distance of 1 meter, and $\alpha_t$ is the BS-RIS path loss exponent. ${\mathbf{f}_{\mathrm{BS}}^{\mathrm{D}}}^{\mathrm{H}}\left(\psi^{\mathrm{D}}\right)$ is the response of the $M$-antenna uniform linear array (ULA) at the BS, based on \cite{wang2019channel}
\begin{align}\label{Rician_model_3}
    {\mathbf{f}_{\mathrm{BS}}^{\mathrm{D}}}^{\mathrm{H}}\left(\psi^{\mathrm{D}}\right)=\left[1,e^{-j2\pi\frac{\delta_a}{\lambda}\sin\psi^{\mathrm{D}}},\cdots,e^{-j2\pi\frac{\delta_a}{\lambda}(M-1)\sin\psi^{\mathrm{D}}}\right],
\end{align}
where $\delta_a$ is the distance between adjacent BS antennas, $\lambda$ is carrier wavelength, $\psi^{\mathrm{D}}$ is the angle of departure (AoD) of signals from the BS. $\mathbf{f}_{\mathrm{RIS}}^{\mathrm{A}}\left(\phi^{\mathrm{A}},\varphi^{\mathrm{A}}\right)$ is the response of an $N=N_x\times N_y$ uniform rectangular planar array (URPA) at the RIS, given by \cite{wang2019channel}
\begin{align}\label{Rician_model_4}
    \notag&\mathbf{f}_{\mathrm{RIS}}^\mathrm{A}\left(\phi^{\mathrm{A}},\varphi^{\mathrm{A}}\right)=\Big[1,\cdots,e^{-j2\pi\frac{\delta_0}{\lambda}\left(n_x\sin\phi^{\mathrm{A}}\cos\varphi^{\mathrm{A}}+n_y\cos\phi^{\mathrm{A}}\right)},\\
    &\quad\cdots,e^{-j2\pi\frac{\delta_0}{\lambda}\left((N_x-1)\sin\phi^{\mathrm{A}}\cos\varphi^{\mathrm{A}}+(N_y-1)\cos\phi^{\mathrm{A}}\right)}\Big]^{\mathrm{T}},
\end{align}
where $0\!\leq\! n_x\!\leq\! N_x-1$, $0\leq\! n_y\!\leq\! N_y-1$, $\delta_0$ is the distance between adjacent RIS elements, $\phi^{\mathrm{A}}$ and $\varphi^{\mathrm{A}}$ are the elevation and azimuth angle of arrival (AoA) of signals to the RIS, respectively.

The NLoS component $\widetilde{\mathbf{G}}\!=\!\left[\widetilde{\mathbf{g}}_{1},\widetilde{\mathbf{g}}_{2},\cdots,\widetilde{\mathbf{g}}_{M}\right]$, $\widetilde{\mathbf{g}}_{m}\in\mathbb{C}^{N\times1}$ is given by
\begin{align}\label{Rician_model_5}
    \widetilde{\mathbf{g}}_{m}\sim\mathcal{CN}\left(\mathbf{0},\varrho_t\mathbf{I}\right), \quad m=1,2,\cdots,M.
\end{align}

The Rician channel model from the RIS to the $k$th user
\begin{align}\label{Rician_model_6}
    \mathbf{h}_{k}^{\mathrm{H}}=\sqrt{\frac{\kappa_{\mathbf{h}_{k}}}{1+\kappa_{\mathbf{h}_{k}}}}\overline{\mathbf{h}}_{k}^{\mathrm{H}}+\sqrt{\frac{1}{1+\kappa_{\mathbf{h}_{k}}}}\widetilde{\mathbf{h}}_{k}^{\mathrm{H}},
\end{align}
where $\kappa_{\mathbf{h}_{k}}$ is the Rician factor, $\overline{\mathbf{h}}^{\mathrm{H}}_k\in\mathbb{C}^{1\times N}$ and $\widetilde{\mathbf{h}}^{\mathrm{H}}_k\in\mathbb{C}^{1\times N}$ represent the LoS and NLoS components, respectively.

The LoS component $\overline{\mathbf{h}}_{k}^{\mathrm{H}}$ is expressed as \cite{wang2019channel}
\begin{align}\label{Rician_model_7}
    \overline{\mathbf{h}}_{k}^{\mathrm{H}}={\mathbf{f}_{\mathrm{RIS}}^{\mathrm{D},k}}^{\mathrm{H}}\left(\phi^{\mathrm{D},k},\varphi^{\mathrm{D},k}\right),
\end{align}
where ${\mathbf{f}_{\mathrm{RIS}}^{\mathrm{D},k}}^{\mathrm{H}}\left(\phi^{\mathrm{D},k},\varphi^{\mathrm{D},k}\right)$ is the response of $N$-element URPA at the RIS, given by \cite{wang2019channel}
\begin{align}\label{Rician_model_8}
    \notag&{\mathbf{f}_{\mathrm{RIS}}^{\mathrm{D},k}}^{\mathrm{H}}\left(\phi^{\mathrm{D},k},\varphi^{\mathrm{D},k}\right)\\
    \notag=&\sqrt{\varrho_{r,k}}\Big[1,\cdots,e^{-j2\pi\frac{\delta_0}{\lambda}\left(n_x\sin\phi^{\mathrm{D},k}\cos\varphi^{\mathrm{D},k}+n_y\cos\phi^{\mathrm{D},k}\right)},\cdots,\\
    &\quad e^{-j2\pi\frac{\delta_0}{\lambda}\left((N_x-1)\sin\phi^{\mathrm{D},k}\cos\varphi^{\mathrm{D},k}+(N_y-1)\cos\phi^{\mathrm{D},k}\right)}\Big],
\end{align}
where $\varrho_{r,k}=C_0d_{r,k}^{-\alpha_r}$ denotes the path loss from the RIS to the $k$th user, in which $d_{r,k}$ denotes the distance between the RIS and the $k$th user, $\alpha_r$ is the RIS-user path loss exponent, and $\phi^{\mathrm{D},k}$ and $\varphi^{\mathrm{D},k}$ are the elevation and azimuth AoD of signals from the RIS to the $k$th user, respectively.

The NLoS component $\widetilde{\mathbf{h}}_{k}$ is given by
\begin{align}\label{Rician_model_9}
    \widetilde{\mathbf{h}}_{k}\sim\mathcal{CN}\left(\mathbf{0},\varrho_{r,k}\mathbf{I}\right), \quad k=1,2,\cdots,K.
\end{align}

In this paper, we assume that instantaneous CSI knowledge can be attained at the BS.

In our RIS-assisted system, the signal is precoded by the TBF at the BS and transmitted to the RIS. The RIS configures the phase shifts of the impinging signals and then reflects them to the users. Therefore, the system model is represented as
\begin{align}\label{channel_model_7}
    \mathbf{y}=\sqrt{P_t}\mathbf{H}\mathbf{\Theta}\mathbf{G}\mathbf{W}\sqrt{\mathbf{\Lambda}}\mathbf{x}+\mathbf{n},
\end{align}
where $\mathbf{x}\in\mathbb{C}^{K\times 1}$ is the transmitted signal vector, $\mathbf{y}\in\mathbb{C}^{K\times 1}$ is the received signal vector, $\mathbf{n}\sim \mathcal{CN}(\mathbf{0},\sigma_n^2\mathbf{I})\in\mathbb{C}^{K\times 1}$ is the circularly symmetric complex Gaussian noise, $\mathbf{W}=[\mathbf{w}_1,\mathbf{w}_2,\cdots,\mathbf{w}_K]\in\mathbb{C}^{M\times K}$ represents the active TBF matrix at the BS, $P_t$ is the total transmitted power of the BS, and $\mathbf{\Lambda}=\mathrm{diag}\{\lambda_1,\lambda_2,\cdots,\lambda_K\}$ is a diagonal power allocation matrix, where $\lambda_k$ represents the power allocated to the signal transmitted to the $k$th user. Hence, in order to normalize the transmit power, we have the following constraints: $\|\mathbf{x}\|=1$, $\|\mathbf{w}_k\|=1$, and $\lambda_1+\lambda_2+\cdots+\lambda_K=1$. Still referring to (\ref{channel_model_7}), $\mathbf{\Theta}$ represents the RIS phase shift matrix, which is diagonal in the conventional RIS architecture, while it is non-diagonal in our proposed RIS architecture. Our objective is to jointly optimize the RIS phase shift matrix $\mathbf{\Theta}$, the TBF matrix $\mathbf{W}$ and the power allocation matrix $\mathbf{\Lambda}$ for maximizing the achievable rate.

In a practical RIS-assisted wireless system as shown in Fig. \ref{Figure_1}, the phase shift matrix $\mathbf{\Theta}$, the TBF matrix $\mathbf{W}$ and the power allocation matrix $\mathbf{\Lambda}$ are jointly optimized at the BS by exploiting the CSI available, i.e. the BS-RIS channel matrix $\mathbf{G}$ and the RIS-users channel matrix $\mathbf{H}$. Then, using the BS-RIS controller link, the optimized phase shift matrix $\mathbf{\Theta}$ is transmitted to the RIS controller, which is responsible for reconfiguring the phase applied to the RIS elements.

In the following, we will briefly highlight the conventional RIS architecture, where the phase shift matrix is diagonal. Then, our proposed non-diagonal RIS architecture will be presented.

\begin{figure}[!t]
\vspace{-0mm}
\setstretch{0.8}
\setlength{\abovecaptionskip}{-3pt}
\setlength{\belowcaptionskip}{-3pt}
    \centering
    \includegraphics[width=3.54in]{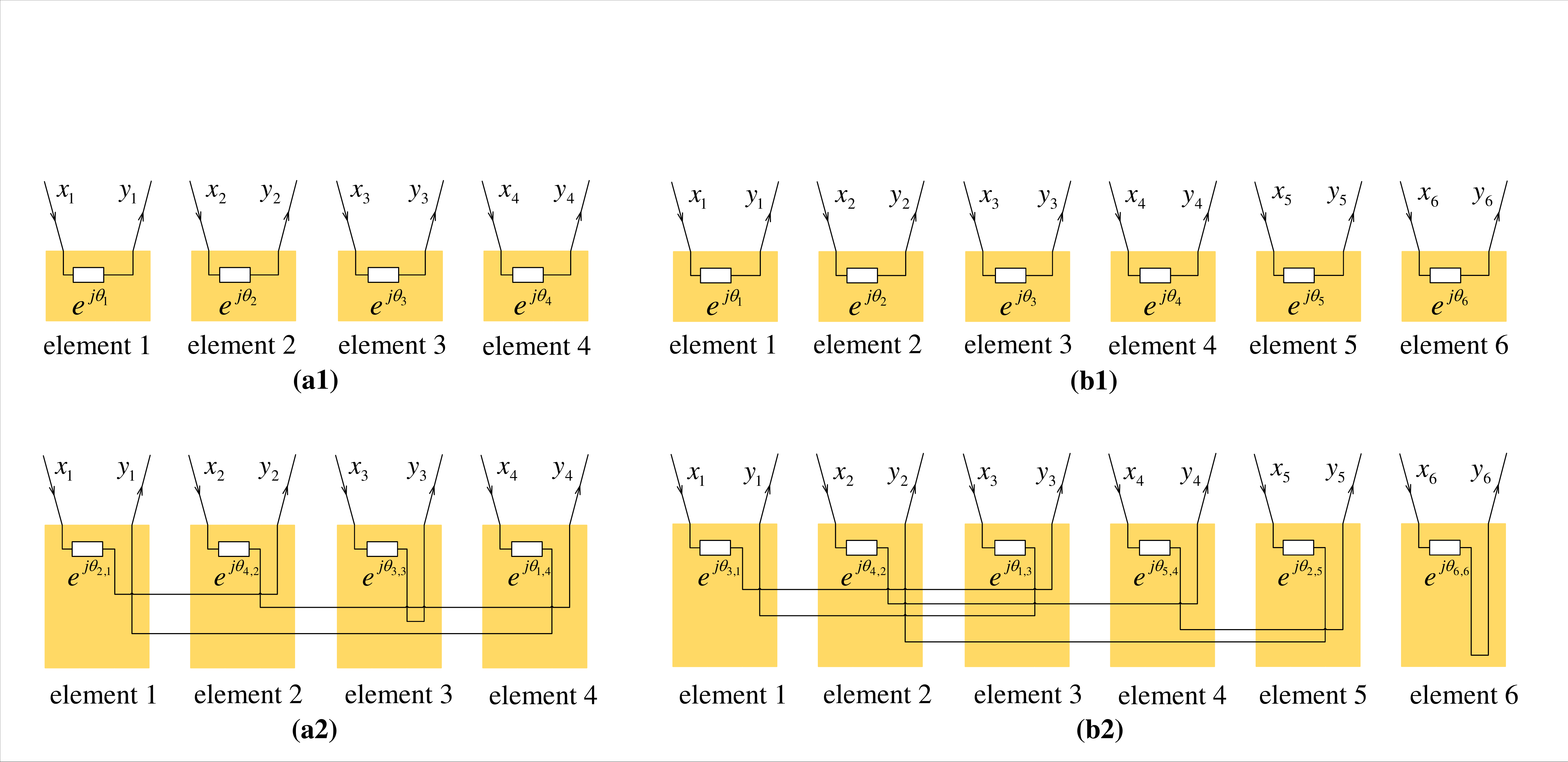}
    \caption{An example of illustration for the conventional RIS architecture having $N=4$ elements and $N=6$ elements in (a1) and (b1) respectively, and the proposed RIS architecture having $N=4$ elements with the bijection function $\mathcal{M}:\{1,2,3,4\}\rightarrow\{2,4,3,1\}$ and $N=6$ elements with the bijection function $\mathcal{M}:\{1,2,3,4,5,6\}\rightarrow\{3,4,1,5,2,6\}$ in (a2) and (b2) respectively.}\label{Figure_2}
\vspace{-5mm}
\end{figure}

\vspace{-5mm}
\subsection{RIS Architecture}
\subsubsection{Conventional RIS Architecture}
Fig. \ref{Figure_2} (a1) shows an example of a conventional 4-element RIS architecture, where each RIS element is `single-connected', i.e. the signal impinging on the $i$th element is only reflected from the $i$th element after phase shift adjustment. Similarly, a conventional 6-element RIS architecture is showed in Fig. \ref{Figure_2} (b1). In conventional RIS-assisted wireless communication systems, each RIS element changes the phase of the impinging signals independently, that is \cite{di2020smart}
\begin{align}\label{channel_model_3}
    y_i=\beta_i x_i e^{j\theta_i},\quad i=1,2,\cdots,N,
\end{align}
where $x_i$ and $y_i$ represent the incident signal and reflected signal of the $i$th RIS element, respectively. The amplitude gain $\beta_i$ is set to 1 to realize full reflection, and the phase shift $\theta_i\in[0,2\pi)$ can be configured for maximizing the channel gain \cite{wu2021intelligent}. Since the phase shift of each RIS element is configured independently, the phase shift matrix, denoted as $\mathbf{\bar{\Theta}}$ \footnote{In this paper, we use $\mathbf{\bar{\Theta}}$ to represent the diagonal phase shift matrix in the conventional RIS architecture, and use $\mathbf{\widetilde{\Theta}}$ to represent the non-diagonal phase shift matrix in our proposed RIS architecture. $\mathbf{\Theta}$ is used when it is not specified which kind of RIS architecture is employed.}, is diagonal and can be represented as
\begin{align}\label{channel_model_4}
    \mathbf{\bar{\Theta}}=\mathrm{diag}\{{e^{j\theta_1},e^{j\theta_2},\cdots,e^{j\theta_N}}\}.
\end{align}
Hence, the equivalent channel spanning from the $m$th BS transmit antenna to the $k$th user can be represented as
\begin{align}\label{channel_model_4_1}
    \notag \mathbf{h}_{k}^{\mathrm{H}}\mathbf{\bar{\Theta}}\mathbf{g}_m=&\sum_{i=1}^{N}[\mathbf{h}_{k}^{\mathrm{H}}]_ie^{j\theta_i}[\mathbf{g}_m]_i\\
    =&\sum_{i=1}^{N}|[\mathbf{h}_{k}^{\mathrm{H}}]_i|\cdot|[\mathbf{g}_m]_i|\cdot e^{j\left(\theta_i+\angle[\mathbf{h}_{k}^{\mathrm{H}}]_i+\angle[\mathbf{g}_m]_i\right)},
\end{align}
which includes the BS-RIS channel, the phase shift applied at the RIS and the RIS-user channel.

On the other hand, if there is a connection between the RIS elements, i.e. the incident signal impinging on the $i$th element can be reflected from other elements, then we will have more flexibility in the design of the RIS phase shift matrix, which can provide an improved performance. The one and only contribution on RIS element cooperation, which was termed as the fully-connected and group-connected RIS architecture, was disseminated by Shen \textit{et al.} \cite{shen2020modeling}. In their solution, the signal impinging on each RIS element was divided into $N$ components, and these $N$ signal components are reflected from $N$ RIS elements after phase shift configuration. This fully connected RIS architecture attains a substantial channel gain. However, its performance enhancement is achieved at the cost of having more entries in the phase shift matrix to optimize, which includes $N\times N$ elements, as opposed to having only $N$ non-zero elements in the conventional design. Furthermore, extra information has to be transmitted over the BS-RIS controller link. On the other hand, the group-connected architecture has significantly lower complexity than the fully-connected architecture, which imposes a modest performance loss. Hence, we propose a novel RIS architecture, which approaches the fully-connected performance at a significantly reduced complexity.

\subsubsection{The proposed RIS Architecture}
Explicitly, we design a novel RIS architecture, where the signal impinging on the $i$th element can be reflected from another one element, denoted as the $i'$th element, after phase shift adjustment. The relationship between the incident signals and the reflected signals can be represented as
\begin{align}\label{channel_model_5}
    y_{i'}=x_{i} e^{j\theta_{i',i}},
\end{align}
where $i$ belongs to the RIS element index set of incident signals $I:\{1,2,\cdots,N\}$, and $i'$ belongs to the RIS element index set of reflected signals $I':\{1,2,\cdots,N\}$. There is a bijective function $\mathcal{M}:I\rightarrow I'$, and $i'=\mathcal{M}(i)$, where the bijection $\mathcal{M}$ is a function between the RIS element indices of incident signals and that of the reflected signals. For example, in Fig. \ref{Figure_2} (a2), we consider an example using four elements, where the signal impinging on the first element is reflected from the second element, thus $\mathcal{M}(1)=2$. Similarly, $\mathcal{M}(2)=4$, $\mathcal{M}(3)=3$, and $\mathcal{M}(4)=1$. In Fig. \ref{Figure_2} (b2), we consider an example using six elements, where the signal impinging on the first element is reflected from the third element, thus $\mathcal{M}(1)=3$. Similarly, $\mathcal{M}(2)=4$, $\mathcal{M}(3)=1$, $\mathcal{M}(4)=5$, $\mathcal{M}(5)=2$, and $\mathcal{M}(6)=6$.

Therefore, in our proposed method the phase shift matrix, denoted as $\mathbf{\widetilde{\Theta}}$, is non-diagonal, and there is only a single non-zero element in each row and each column. The phase shift matrix in Fig. \ref{Figure_2} (a2) can be represented as
\begin{align}\label{channel_model_6}
    \mathbf{\widetilde{\Theta}}=
        \left[\begin{array}{cccc}
             0 & 0 & 0 & e^{j\theta_{1,4}}\\
             e^{j\theta_{2,1}} & 0 & 0 & 0\\
             0 & 0 & e^{j\theta_{3,3}} & 0\\
             0 & e^{j\theta_{4,2}} & 0 & 0
         \end{array}\right].
\end{align}
Similarly, the phase shift matrix in Fig. \ref{Figure_2} (b2) can be represented as
\begin{align}\label{channel_model_6_1}
    \mathbf{\widetilde{\Theta}}=
        \left[\begin{array}{cccccccc}
             0 & 0 & e^{j\theta_{1,3}} & 0 & 0 & 0\\
             0 & 0 & 0 & 0 & e^{j\theta_{2,5}} & 0\\
             e^{j\theta_{3,1}} & 0 & 0 & 0 & 0 & 0\\
             0 & e^{j\theta_{4,2}} & 0 & 0 & 0 & 0\\
             0 & 0 & 0 & e^{j\theta_{5,4}} & 0 & 0\\
             0 & 0 & 0 & 0 & 0 & e^{j\theta_{6,6}}
         \end{array}\right].
\end{align}
Since only $N$ non-zero entries of the phase shift matrix have to be optimized and $N$ position information values of these non-zero entries have to be recorded in each coherence time for our RIS-aided systems, this only modestly increases the optimisation complexity and the amount of information transmitted over the BS-RIS controller link, compared to the conventional architecture.

Since the phase shift matrix in our proposed method is non-diagonal, we may refer to it as the \textit{RIS architecture with non-diagonal phase shift matrix}. Compared to the conventional RIS architecture with diagonal phase shift matrix, the proposed non-diagonal phase shift matrix method has the potential of attaching higher channel gain. Let us consider an example using a 4-element RIS employed in a SISO system, and assume that the channel vector of the link from the BS to the RIS is $\mathbf{g}\!=\!\left[1.4e^{-j\frac{3\pi}{4}},0.2e^{j\frac{5\pi}{6}},0.4e^{-j\frac{7\pi}{8}},0.8e^{-j\frac{\pi}{6}}\right]^\mathrm{T}$, while that of the link from the RIS to the user is $\mathbf{h}^\mathrm{H}\!=\!\left[0.6e^{-j\frac{\pi}{4}},1e^{j\frac{2\pi}{3}},0.3e^{j\frac{\pi}{3}},0.1e^{j\frac{\pi}{8}}\right]$. In the conventional phase shift matrix based method, the channel gain can be maximized when the RIS phase shifts are designed coherently, i.e., $\theta_1\!=\!-\left(\angle[\mathbf{g}]_1+\angle[\mathbf{h}^\mathrm{H}]_1\right)\!=\!\pi$, $\theta_2\!=\!-\left(\angle[\mathbf{g}]_2+\angle[\mathbf{h}^\mathrm{H}]_2\right)\!=\!-\frac{3\pi}{2}$, $\theta_3\!=\!-\left(\angle[\mathbf{g}]_3+\angle[\mathbf{h}^\mathrm{H}]_3\right)\!=\!\frac{5\pi}{8}$, $\theta_4\!=\!-\left(\angle[\mathbf{g}]_4+\angle[\mathbf{h}^\mathrm{H}]_4\right)\!=\!\frac{\pi}{24}$. Then, the corresponding channel gain is given by $1.4\times0.6+0.2\times1+0.4\times0.3+0.8\times0.1=12.4$. By contrast, in our proposed non-diagonal phase shift matrix method, if the phase shift matrix is designed as the structure in (\ref{channel_model_6}), i.e. when the bijective function is $\mathcal{M}(1)=2$, $\mathcal{M}(2)=4$, $\mathcal{M}(3)=3$, $\mathcal{M}(4)=1$, and the RIS phase shifts are designed coherently, i.e., $\theta_{1,4}\!=\!-\left(\angle[\mathbf{g}]_4+\angle[\mathbf{h}^\mathrm{H}]_1\right)\!=\!\frac{5\pi}{12}$, $\theta_{2,1}\!=\!-\left(\angle[\mathbf{g}]_1+\angle[\mathbf{h}^\mathrm{H}]_2\right)\!=\!\frac{\pi}{12}$, $\theta_{3,3}\!=\!-\left(\angle[\mathbf{g}]_3+\angle[\mathbf{h}^\mathrm{H}]_3\right)\!=\!\frac{13\pi}{24}$, $\theta_{4,2}\!=\!-\left(\angle[\mathbf{g}]_2+\angle[\mathbf{h}^\mathrm{H}]_4\right)\!=\!-\frac{23\pi}{24}$, then the corresponding channel gain is given by $0.8\times0.6+1.4\times1+0.4\times0.3+0.2\times0.1=20.2$. Therefore, higher channel gain can be achieved when the bijective function $\mathcal{M}$ and the RIS phase shifts of the non-diagonal phase shift matrix are appropriately designed. The details of optimizing our proposed non-diagonal RIS phase shift matrix will be discussed in Section \ref{Section_Beamforming_Design}, where the bijection function $\mathcal{M}$ and the values of each element's phase shift can be obtained in the non-diagonal phase shift matrix $\mathbf{\widetilde{\Theta}}$.

\begin{figure}[!t]
\vspace{-0mm}
\setstretch{0.8}
\setlength{\abovecaptionskip}{-3pt}
\setlength{\belowcaptionskip}{-3pt}
    \centering
    \includegraphics[width=3.5in]{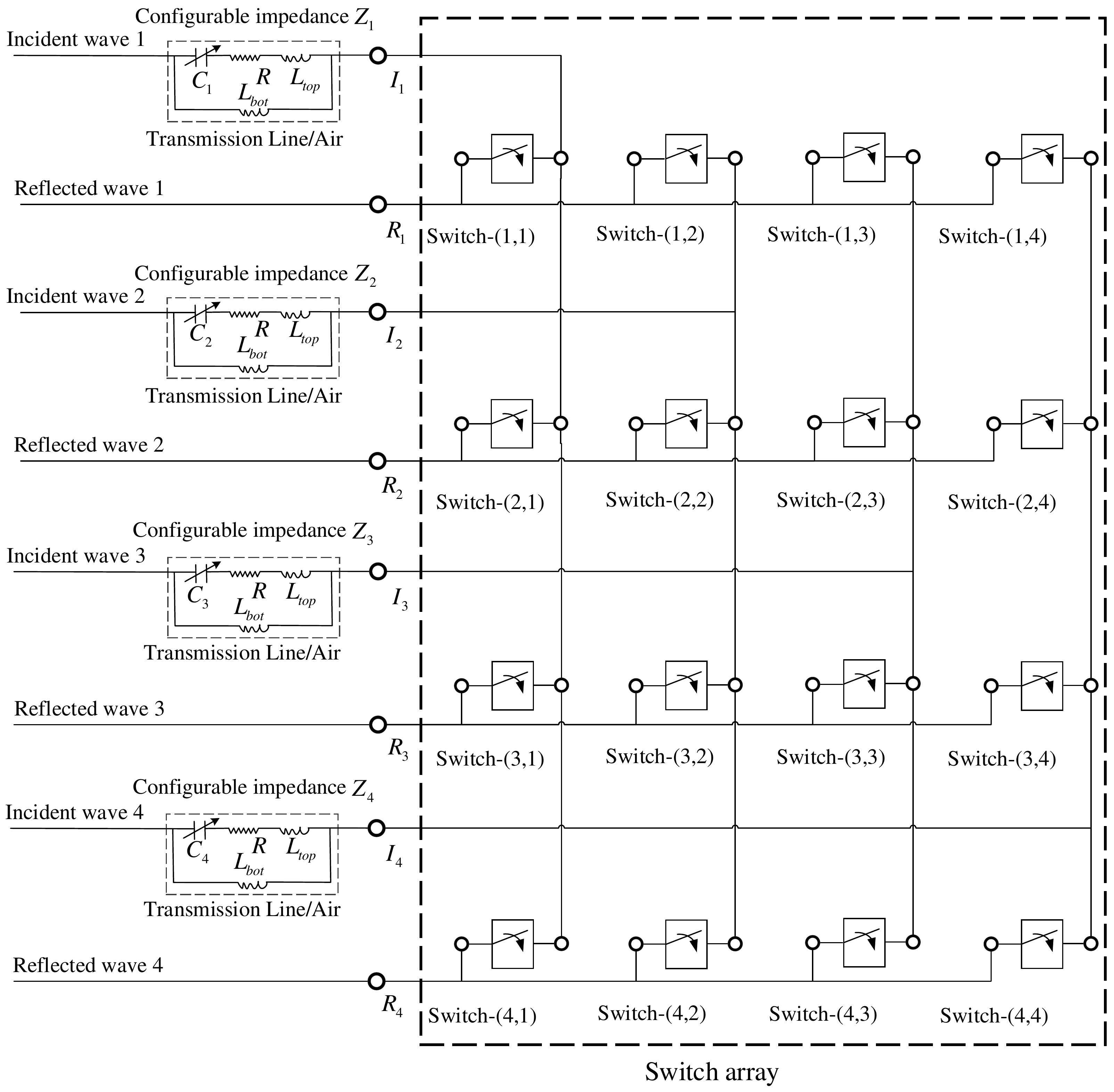}
    \caption{The transmission line model of our proposed non-diagonal phase shift matrix architecture with $N=4$ RIS elements.}\label{Figure_3}
\vspace{-5mm}
\end{figure}

\vspace{-5mm}
\subsection{Implementation Circuits}
Although the implementational specifics of our proposed RIS architecture are beyond the scope of this paper, in the following we briefly highlight a potential implementation suitable for our proposed architecture.

In \cite{shen2020modeling}, the authors employed scattering parameter network models based on reciprocal architectures for describing the implementation of the conventional RIS structure and the fully-connected/group-connected RIS structure. The corresponding phase shift matrices are symmetric in \cite{shen2020modeling}. On the other hand, our proposed architecture relies on a potentially non-symmetric matrix structure due to the fact that our design requires non-reciprocal connections.

In the following, we employ the classic transmission line model \cite{abeywickrama2020intelligent} for highlighting design the implementation of our proposed RIS architecture. The circuits of the conventional RIS architecture have been presented in \cite{abeywickrama2020intelligent,liu2021reconfigurable}. When it comes to using the transmission line model based design of our RIS architecture, we can employ switch arrays for connecting the different RIS elements. Fig. \ref{Figure_3} illustrates an example of the transmission line model of our RIS architecture with $N=4$ elements. In each RIS element, the reconfigurable impedance includes a bottom layer inductance $L_{bot}$, a top layer inductance $L_{top}$, an effective resistance $R$, and a variable capacitance $C_n$, where $n=1,2,\cdots,N$ \cite{liu2021reconfigurable}. The phase shift of the reconfigurable impedance $Z_n$ is controlled by its variable capacitance $C_n$. To realize an $N$-element non-diagonal phase shift matrix, an array of $N\times N$ switches is required. The ON/OFF state of these switches is determined by the positions of non-zero entries in the RIS phase shift matrix. Specifically, the switches are turned on if the corresponding element in the RIS phase shift matrix is non zero, while they are turned off, if the corresponding elements are zero. For example, to realize the non-diagonal phase shift matrix of (\ref{channel_model_6}), Switch-(2,1), Switch-(4,2), Switch-(3,3) and Switch-(1,4) are turned on, while the other switches are turned off in Fig. \ref{Figure_3}. In this case, the signal impinging on the first RIS element is reflected from the second RIS element after phase shift configuration, while the signal impinging on the second RIS element is reflected from the fourth RIS element after phase shift configuration, etc. A potential implementation for the switches relies on using RF micro-electromechanical systems (MEMS) \cite{rebeiz2001rf}, which have been widely used in wireless communication systems as a benefit of their near-zero power consumption, high isolation, low insertion loss, low intermodulation products and low cost.

There are also other potential implementations for our proposed architecture. Explicitly, since we have to route the signal between the different elements depending on the bijection function $\mathcal{M}$, radio frequency couplers and isolators can be employed \cite{laur2021c,yang2015study,wang2011self,d2004substrate}. Additionally, passive phase shifters way also be employed in conjunction with these couplers for attaining accurate phase shifts \cite{adhikari2013tunable}. Finally, the authors of \cite{li2018metasurfaces}, \cite{liaskos2020internet} provided comprehensive discussions of metasurfaces, including their operation and functionalities. Hence the proposed architecture is viable and has several existing implementations, which can be further optimized in our future researches.

\section{Beamforming Design}\label{Section_Beamforming_Design}
In the previous section, we presented a novel RIS architecture, while here we present our joint beamforming design maximizing the attainable rate. This is achieved by jointly optimizing the passive beamforming matrix of the RIS and the TBF matrix of the BS. We start with deriving the closed-form solution of the SISO case, and then extend our analysis to more general single-user MISO case and multi-user MIMO case by employing the alternating optimization method and semi-definite relaxation technique \cite{boyd2004convex}, respectively.

\vspace{-5mm}
\subsection{Beamforming Design for SISO Systems}
In SISO systems, both the BS and the single user are equipped with a single antenna, while the RIS has $N$ reflecting elements. The system model in (\ref{channel_model_7}) can be written as
\begin{align}\label{beamforming_design_1}
    y=\sqrt{P_t}\mathbf{h}^{\mathrm{H}}\mathbf{\Theta}\mathbf{g}x+n,
\end{align}
where $x\in\mathbb{C}^{1\times 1}$ is the transmitted signal, $\mathbf{g}\in\mathbb{C}^{N\times 1}$ is the channel vector arriving from the BS to the RIS, $\mathbf{h}^{\mathrm{H}}\in\mathbb{C}^{1\times N}$ is the channel vector of the link spanning from the RIS to the user, and $n\in\mathbb{C}^{1\times 1}$ is the circularly symmetric complex Gaussian noise. The achievable rate of this SISO link is given by
\begin{align}\label{beamforming_design_1_1}
    R_{\mathrm{SISO}}=\log_2\left(1+\frac{P_t}{\sigma_n^2}|\mathbf{h}^{\mathrm{H}}\mathbf{\Theta}\mathbf{g}|^2\right).
\end{align}
Our aim is to find the phase shift matrix $\mathbf{\Theta}$ that maximizes the rate, which can be formulated as
\begin{align}\label{beamforming_design_1_2}
    \mathrm{(P1.a)}\qquad&\max_{\mathbf{\Theta}}\ \log_2\left(1+\frac{P_t}{\sigma_n^2}|\mathbf{h}^{\mathrm{H}}\mathbf{\Theta}\mathbf{g}|^2\right)\\
    \notag \mathrm{s.t.}&\quad 0\leq\theta_{\mathcal{M}(i),i}<2\pi,\quad i=1,2,\cdots,N.
\end{align}
Similar to \cite{wu2021intelligent}, by ignoring the constant terms, the achievable rate optimization problem of SISO systems is equivalent to maximizing the channel gain as follows
\begin{align}\label{beamforming_design_3}
    \mathrm{(P1.b)}\qquad&\max_{\mathbf{\Theta}}|\mathbf{h}^{\mathrm{H}}\mathbf{\Theta}\mathbf{g}|^2\\
    \notag \mathrm{s.t.}&\quad 0\leq\theta_{\mathcal{M}(i),i}<2\pi,\quad i=1,2,\cdots,N.
\end{align}

\subsubsection{Conventional RIS Architecture}
Again, in the conventional RIS architecture, the phase shift matrix is diagonal, so the bijection $f:I\rightarrow I'$ is essentially $\mathcal{M}(i)=i$. Therefore, the channel gain $|\mathbf{h}^{\mathrm{H}}\mathbf{\bar{\Theta}}\mathbf{g}|^2$ is given by
\begin{align}\label{beamforming_design_4}
    |\mathbf{h}^{\mathrm{H}}\mathbf{\bar{\Theta}}\mathbf{g}|^2=\left(\sum_{i=1}^{N}[\mathbf{h}^{\mathrm{H}}]_i[\mathbf{g}]_ie^{j\theta_i}\right)^2,
\end{align}
where the optimal solution for $\mathbf{\bar{\Theta}}$ can be obtained as \cite{wu2021intelligent}
\begin{align}\label{beamforming_design_5}
    \theta_i=-(\angle[\mathbf{h}^{\mathrm{H}}]_i+\angle[\mathbf{g}]_i), \quad i=1,2,\cdots,N,
\end{align}
which essentially aligns all the signals reflected by the RIS with the impinging signals to arrange for their coherent combination, and the maximum channel gain based on (\ref{beamforming_design_5}) can be expressed as \cite{wu2021intelligent}
\begin{align}\label{beamforming_design_6}
    \max|\mathbf{h}^{\mathrm{H}}\mathbf{\bar{\Theta}}\mathbf{g}|^2=\varrho_t\varrho_r\left(\sum_{i=1}^{N}a_ib_i\right)^2,
\end{align}
where $a_i$ and $b_i$ are the amplitude of $\frac{1}{\sqrt{\varrho_t}}[\mathbf{g}]_i$ and $\frac{1}{\sqrt{\varrho_r}}[\mathbf{h}^{\mathrm{H}}]_i$, respectively. Observe from (\ref{beamforming_design_6}) that in conventional RIS architectures, the maximum channel gain is proportional to the square of the sum of $a_ib_i$, in which $a_i$ and $b_i$ are amalgamated based on the Equal Gain Combining (EGC) criterion.

\subsubsection{The proposed RIS Architecture}
Since in our proposed RIS architecture the signal impinging on the $i$th element can be reflected from the $i'$th element after phase shift adjustment, the channel gain can be written as
\begin{align}\label{beamforming_design_7}
    |\mathbf{h}^{\mathrm{H}}\mathbf{\widetilde{\Theta}}\mathbf{g}|^2=\left(\sum_{i=1}^{N}[\mathbf{h}^{\mathrm{H}}]_{i'}[\mathbf{g}]_ie^{j\theta_{i',i}}\right)^2=\varrho_t\varrho_r\left(\sum_{i=1}^{N}a_ib_{i'}\right)^2,
\end{align}
where
\begin{align}\label{beamforming_design_8}
    \theta_{i',i}=-(\angle[\mathbf{h}^{\mathrm{H}}]_{i'}+\angle[\mathbf{g}]_i), \quad i=1,2,\cdots,N,\quad i'=\mathcal{M}(i).
\end{align}
Hence, first we should aim for finding the function $\mathcal{M}:I\rightarrow I'$ to maximize the channel gain in (\ref{beamforming_design_7}). According to the Maximum Ratio Combining (MRC) criterion, the maximum of the channel gain in (\ref{beamforming_design_7}) is given by
\begin{align}\label{beamforming_design_9}
    \max|\mathbf{h}^{\mathrm{H}}\mathbf{\widetilde{\Theta}}\mathbf{g}|^2=\varrho_t\varrho_r\left(\sum_{i=1}^{N}a_{(i)}b_{(i)}\right)^2,
\end{align}
where $a_{(1)},a_{(2)},\cdots,a_{(N)}$ represents the sequence of $a_1,a_2,\cdots,a_N$ sorted in an ascending order, and similarly, $b_{(1)},b_{(2)},\cdots,b_{(N)}$ is the sequence of $b_1,b_2,\cdots,b_N$ sorted in an ascending order. According to \cite{zhang2017matrix}, when a permutation matrix is multiplied from the left by a column vector, it will permute the elements of the column vector, while when a permutation matrix is multiplied from the right by a row vector, it will permute the elements of the row vector. Therefore, the channel gain in (\ref{beamforming_design_9}) can be written as
\begin{align}\label{beamforming_design_10}
    \max|\mathbf{h}^{\mathrm{H}}\mathbf{\widetilde{\Theta}}\mathbf{g}|^2=|\mathbf{h}^{\mathrm{H}}\mathbf{J}_r\mathbf{\bar{\Theta}}\mathbf{J}_t\mathbf{g}|^2,
\end{align}
where $\mathbf{J}_t$ is a permutation matrix, which sorts the element amplitude in the column vector $\mathbf{g}$ in an ascending order, yielding $\frac{1}{\sqrt{\varrho_t}}|\mathbf{J}_t\mathbf{g}|=[a_{(1)},a_{(2)},\cdots,a_{N}]^{\mathrm{T}}$. Still referring to (\ref{beamforming_design_10}), $\mathbf{J}_r$ is a permutation matrix, which sorts the element amplitude in the row vector $\mathbf{h}^{\mathrm{H}}$ in an ascending order, leading to $\frac{1}{\sqrt{\varrho_r}}|\mathbf{h}^{\mathrm{H}}\mathbf{J}_r|=[b_{(1)},b_{(2)},\cdots,b_{(N)}]$. In this case, the channel vectors $\mathbf{g}$ and $\mathbf{h}^{\mathrm{H}}$ can be combined based on the MRC criterion. Furthermore, $\mathbf{\bar{\Theta}}$ in (\ref{beamforming_design_10}) is a diagonal phase shift matrix, in which the $i$th diagonal element is given by
\begin{align}\label{beamforming_design_11}
    \theta_i=-(\angle[\mathbf{h}^{\mathrm{H}}\mathbf{J}_r]_i+\angle[\mathbf{J}_t\mathbf{g}]_i),\quad i=1,2,\cdots,N.
\end{align}
Therefore, the optimal non-diagonal phase matrix is formulated as $\mathbf{\widetilde{\Theta}}_{\mathrm{opt}}=\mathbf{J}_r\mathbf{\bar{\Theta}}\mathbf{J}_t$.

\begin{figure}[!t]
\vspace{-0mm}
\setstretch{0.8}
\setlength{\abovecaptionskip}{-3pt}
\setlength{\belowcaptionskip}{-3pt}
    \centering
    \includegraphics[width=2.2in]{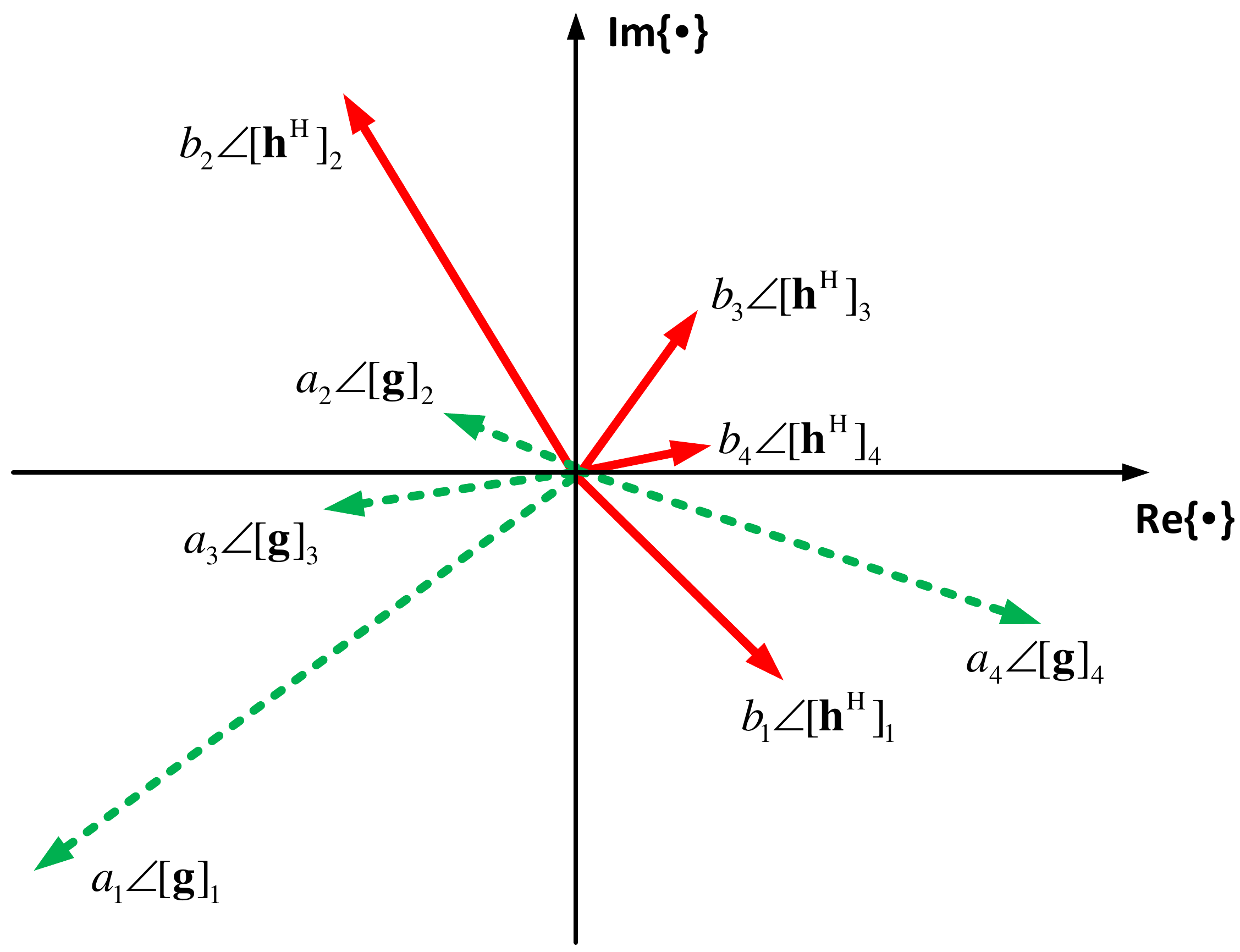}
    \caption{The illustration of channel vectors in the example of a 4-element RIS-assisted system.}\label{Figure_4}
\vspace{-5mm}
\end{figure}

To expound further, in Fig. \ref{Figure_4}, we present an example of a 4-element RIS-assisted system, where the '$\dashrightarrow$' vectors represent the BS-RIS channel vector $\mathbf{g}$, while the '$\rightarrow$' vectors correspond to the RIS-user channel vector $\mathbf{h}^{\mathrm{H}}$. It can be observed that $a_2\!<\!a_3\!<\!a_4\!<\!a_1$ and $b_4\!<\!b_3\!<\!b_1\!<\!b_2$, so $a_{(1)}\!=\!a_2,a_{(2)}\!=\!a_3,a_{(3)}\!=\!a_4,a_{(4)}\!=\!a_1$, and $b_{(1)}\!=\!b_4,b_{(2)}\!=\!b_3,b_{(3)}\!=\!b_1,b_{(4)}\!=\!b_2$. Therefore, in our proposed architecture, the permutation matrices $\mathbf{J}_t$ and $\mathbf{J}_r$ are derived as
\begin{align}\label{permutation_matrix_J}
    \mathbf{\mathbf{J}}_t=
        \left[\begin{array}{cccc}
             0 & 1 & 0 & 0\\
             0 & 0 & 1 & 0\\
             0 & 0 & 0 & 1\\
             1 & 0 & 0 & 0
         \end{array}\right],\quad
    \mathbf{\mathbf{J}}_r=
        \left[\begin{array}{cccc}
             0 & 0 & 1 & 0\\
             0 & 0 & 0 & 1\\
             0 & 1 & 0 & 0\\
             1 & 0 & 0 & 0
         \end{array}\right],
\end{align}
and the bijective function is given by $\mathcal{M}(1)=2,\mathcal{M}(2)=4,\mathcal{M}(3)=3,\mathcal{M}(4)=1$. Hence, the non-diagonal phase shift matrix $\mathbf{\widetilde{\Theta}}$ is designed as
\begin{align}\label{Theta_example}
    \notag \mathbf{\widetilde{\Theta}}&=\mathbf{J}_r\mathbf{\bar{\Theta}}\mathbf{J}_t\\
    \notag &=\mathbf{J}_r\cdot\mathrm{diag}\{{e^{j\theta_1},e^{j\theta_2},\cdots,e^{j\theta_N}}\}\cdot\mathbf{J}_t\\
    &=\left[\begin{array}{cccc}
             0 & 0 & 0 & e^{j\theta_3}\\
             e^{j\theta_4} & 0 & 0 & 0\\
             0 & 0 & e^{j\theta_2} & 0\\
             0 & e^{j\theta_1} & 0 & 0
         \end{array}\right],
\end{align}
where $e^{j\theta_1}=e^{j\theta_{4,2}}=-(\angle[\mathbf{h}^{\mathrm{H}}]_4+\angle[\mathbf{g}]_2)$, $e^{j\theta_2}=e^{j\theta_{3,3}}=-(\angle[\mathbf{h}^{\mathrm{H}}]_3+\angle[\mathbf{g}]_3)$, $e^{j\theta_3}=e^{j\theta_{1,4}}=-(\angle[\mathbf{h}^{\mathrm{H}}]_1+\angle[\mathbf{g}]_4)$, $e^{j\theta_4}=e^{j\theta_{2,1}}=-(\angle[\mathbf{h}^{\mathrm{H}}]_2+\angle[\mathbf{g}]_1)$. Then, the corresponding channel gain is optimized as
\begin{align}\label{channel_gain_example}
    \notag |\mathbf{h}^{\mathrm{H}}\mathbf{\widetilde{\Theta}}\mathbf{g}|^2&=\varrho_t\varrho_r\left(a_{(1)}b_{(1)}+a_{(2)}b_{(2)}+a_{(3)}b_{(3)}+a_{(4)}b_{(4)}\right)^2\\
    &=\varrho_t\varrho_r\left(a_2b_4+a_3b_3+a_4b_1+a_1b_2\right)^2.
\end{align}

We observe from (\ref{beamforming_design_9}) that in our proposed RIS architecture, the maximum channel gain is proportional to the square of the sum of $a_{(i)}b_{(i)}$, in which $a_{(i)}$ and $b_{(i)}$ are combined by obeying the MRC criterion, when the number of RIS elements $N$ is large.

Finally, when the optimal phase shift matrix $\mathbf{\widetilde{\Theta}}_{\mathrm{opt}}$ is attained, the achievable rate is given by
\begin{align}\label{beamforming_design_12}
    R_{\mathrm{SISO}}=\log_2\left(1+\frac{P_t}{\sigma_n^2}|\mathbf{h}^{\mathrm{H}}\mathbf{\widetilde{\Theta}}_{\mathrm{opt}}\mathbf{g}|^2\right).
\end{align}

\vspace{-4mm}
\subsection{Beamforming Design for Single-user MISO Systems}
In single-user MISO systems, the BS is equipped with $M$ downlink transmit antennas and the single user is equipped with a single receiver antenna, while the RIS has $N$ elements. The system model in (\ref{channel_model_7}) can be written as
\begin{align}\label{beamforming_design_miso_1}
    y=\sqrt{P_t}\mathbf{h}^{\mathrm{H}}\mathbf{\widetilde{\Theta}}\mathbf{G}\mathbf{w}x+n,
\end{align}
and the achievable rate is given by
\begin{align}\label{beamforming_design_miso_1_1}
    R_{\mathrm{MISO}}=\log_2\left(1+\frac{P_t}{\sigma_n^2}|\mathbf{h}^{\mathrm{H}}\mathbf{\widetilde{\Theta}}\mathbf{G}\mathbf{w}|^2\right).
\end{align}
The problem of maximizing the achievable rate can be formulated as
\begin{align}\label{beamforming_design_miso_1_2}
    \mathrm{(P2.a)}\qquad&\max_{\mathbf{\widetilde{\Theta}},\mathbf{w}}\ \log_2\left(1+\frac{P_t}{\sigma_n^2}|\mathbf{h}^{\mathrm{H}}\mathbf{\widetilde{\Theta}}\mathbf{G}\mathbf{w}|^2\right)\\
    \notag \mathrm{s.t.}&\quad \|\mathbf{w}\|=1,\\
    \notag &\quad 0\leq\theta_{\mathcal{M}(i),i}<2\pi,\quad i=1,2,\cdots,N.
\end{align}
Similar to SISO cases, the achievable rate optimization problem of single-user MISO systems is equivalent to maximizing the channel gain as follows
\begin{align}\label{beamforming_design_miso_3}
    \mathrm{(P2.b)}\qquad&\max_{\mathbf{\widetilde{\Theta}},\mathbf{w}}|\mathbf{h}^{\mathrm{H}}\mathbf{\widetilde{\Theta}}\mathbf{G}\mathbf{w}|^2\\
    \notag \mathrm{s.t.}&\quad \|\mathbf{w}\|=1,\\
    \notag &\quad 0\leq\theta_{\mathcal{M}(i),i}<2\pi,\quad i=1,2,\cdots,N.
\end{align}
Since (P2.b) represents a non-convex problem, we employ the popular alternating optimization method for solving it iteratively.

Firstly, when the TBF vector $\mathbf{w}$ is given, $\mathbf{Gw}$ becomes a column vector, and the phase shift matrix $\mathbf{\widetilde{\Theta}}$ can be designed similarly as in the SISO case.

Secondly, when the phase shift matrix $\mathbf{\widetilde{\Theta}}$ is given, the equivalent channel can be obtained as $\mathbf{h}^\mathrm{H}\mathbf{\widetilde{\Theta}}\mathbf{G}$. Then the TBF vector can be designed based on the maximum ratio transmission (MRT) method, yielding $\mathbf{w}=\frac{(\mathbf{h}^\mathrm{H}\mathbf{\widetilde{\Theta}}\mathbf{G})^\mathrm{H}}{\|\mathbf{h}^\mathrm{H}\mathbf{\widetilde{\Theta}}\mathbf{G}\|}$. The detailed process of the alternating optimization method conceived for RIS-assisted single-user MISO systems is shown in Algorithm \ref{algorithm_1}.

When the optimal phase shift matrix $\mathbf{\widetilde{\Theta}}_{\mathrm{opt}}$ and TBF vector $\mathbf{w}_{\mathrm{opt}}$ are obtained, the achievable rate is given by
\begin{align}\label{beamforming_design_miso_3_1}
    R_{\mathrm{MISO}}=\log_2\left(1+\frac{P_t}{\sigma_n^2}|\mathbf{h}^{\mathrm{H}}\mathbf{\widetilde{\Theta}}_{\mathrm{opt}}\mathbf{G}\mathbf{w}_{\mathrm{opt}}|^2\right).
\end{align}

\begin{algorithm}
\setstretch{0}
\caption{Alternating optimization method for RIS-assisted single-user MISO systems}
\label{algorithm_1}
\begin{algorithmic}[1]
\REQUIRE
    BS-RIS channel matrix $\mathbf{G}$, and RIS-user channel vector $\mathbf{h}^\mathrm{H}$.
\ENSURE
    The optimal phase shift matrix $\mathbf{\widetilde{\Theta}}_{\mathrm{opt}}$, and the optimal transmission beamforming vector $\mathbf{w}_{\mathrm{opt}}$.
    \STATE
        Choose the proper permutation matrix $\mathbf{J}_r$ which sorts the element amplitude of row vector $\mathbf{h}^\mathrm{H}$ in an ascending order.
    \STATE
        Set an initial $\mathbf{w}$ satisfying $\|\mathbf{w}\|=1$.
    \STATE
        \textbf{Repeat}
    \STATE
        \qquad Choose the proper permutation matrix $\mathbf{J}_t$ which can sort the element amplitude of column vector $\mathbf{Gw}$ in an ascending order.
    \STATE
        \qquad According to the row vector $\mathbf{h}^\mathrm{H}\mathbf{J}_r$ and column vector $\mathbf{J}_t\mathbf{Gw}$, the elements in diagonal matrix $\mathbf{\bar{\Theta}}$ is designed as $\theta_i=-\left(\angle{[\mathbf{h}^\mathrm{H}\mathbf{J}_r]_i}+\angle{[\mathbf{J}_t\mathbf{Gw}]_i}\right)$, $i=1,2,\cdots,N$.
    \STATE
        \qquad Design the transmission beamforming vector based on MRT criterion as $\mathbf{w}=\frac{(\mathbf{h}^\mathrm{H}\mathbf{J}_r\mathbf{\bar{\Theta}}\mathbf{J}_t\mathbf{G})^\mathrm{H}}{\|\mathbf{h}^\mathrm{H}\mathbf{J}_r\mathbf{\bar{\Theta}}\mathbf{J}_t\mathbf{G}\|}$.
    \STATE
        \textbf{Until} reaching the maximal number of iterations or the increment of the objective value is smaller than threshold $\epsilon$.
    \STATE
        \textbf{Return} the optimal phase shift matrix $\mathbf{\widetilde{\Theta}}_{\mathrm{opt}}=\mathbf{J}_r\mathbf{\bar{\Theta}}\mathbf{J}_t$ and the optimal transmission beamforming vector $\mathbf{w}_{\mathrm{opt}}=\frac{(\mathbf{h}^\mathrm{H}\mathbf{\widetilde{\Theta}}\mathbf{G})^\mathrm{H}}{\|\mathbf{h}^\mathrm{H}\mathbf{\widetilde{\Theta}}\mathbf{G}\|}$.
\end{algorithmic}
\end{algorithm}

\vspace{-5mm}
\subsection{Beamforming Design for Multi-user MIMO Systems}
In multi-user MIMO systems, including a BS having $M$ transmit antennas and $K$ single-antenna users, the system model is given by
\begin{align}\label{beamforming_design_mimo_A0}
    \mathbf{y}=\sqrt{P_t}\mathbf{H}\mathbf{\widetilde{\Theta}}\mathbf{G}\mathbf{W}\sqrt{\mathbf{\Lambda}}\mathbf{x}+\mathbf{n},
\end{align}
and the achievable rate is formulated as
\begin{align}\label{beamforming_design_mimo_A1}
    R_{\mathrm{MIMO}}=\log_2\left|\mathbf{I}+\frac{P_t}{\sigma_n^2}\mathbf{H}\mathbf{\widetilde{\Theta}}\mathbf{G}\mathbf{W}\mathbf{\Lambda}\mathbf{W}^{\mathrm{H}}\mathbf{G}^{\mathrm{H}}\mathbf{\widetilde{\Theta}}^{\mathrm{H}}\mathbf{H}^{\mathrm{H}}\right|.
\end{align}
The problem of maximizing the achievable rate can be formulated as
\begin{align}\label{beamforming_design_mimo_A2}
    \mathrm{(P3.a)}\ &\max_{\mathbf{W},\mathbf{\widetilde{\Theta}},\mathbf{\Lambda}}\log_2\!\left|\mathbf{I}\!+\!\frac{P_t}{\sigma_n^2}\mathbf{H}\mathbf{\widetilde{\Theta}}\mathbf{G}\mathbf{W}\mathbf{\Lambda}\mathbf{W}^{\mathrm{H}}\mathbf{G}^{\mathrm{H}}\mathbf{\widetilde{\Theta}}^{\mathrm{H}}\mathbf{H}^{\mathrm{H}}\right|\!\\
    \notag \mathrm{s.t.}&\quad \|\mathbf{w}_k\|=1,\ k=1,2,\cdots,K,\\
    \notag &\quad \lambda_1+\lambda_2+\cdots+\lambda_K=1,\\
    \notag &\quad 0\leq\theta_{\mathcal{M}(i),i}<2\pi,\quad i=1,2,\cdots,N.
\end{align}
According to \cite{wu2019intelligent}, the two-stage algorithm, which decouples the joint beamforming design problem (P3.a) into two subproblems, has lower computational complexity, while suffering from a slight performance erosion, when compared to alternating optimization algorithm. Therefore, we employ the two-stage algorithm for maximizing the achievable rate of our proposed RIS architecture as follows.
\begin{itemize}
\item \textit{Stage I:} The RIS phase shift matrix is optimized by maximizing the sum of the combined channel gain of all users, which can be expressed as
    \begin{align}\label{beamforming_design_mimo_A2_1}
        \mathrm{(P3.b1)}\quad&\max_{\mathbf{\widetilde{\Theta}}}\ \sum_{k=1}^{K}\|\mathbf{h}_k^{\mathrm{H}}\mathbf{\widetilde{\Theta}}\mathbf{G}\|^2\\
        \notag \mathrm{s.t.}&\quad 0\leq\theta_{\mathcal{M}(i),i}<2\pi,\quad i=1,2,\cdots,N.
    \end{align}

    We employ the SDR method for solving the problem (P3.b1). Specifically, let us define a column vector $\mathbf{q}=[e^{j\theta_1},e^{j\theta_2},\cdots,e^{j\theta_N}]^{\mathrm{H}}$. Then $\sum_{k=1}^{K}\|\mathbf{h}_k^{\mathrm{H}}\mathbf{\widetilde{\Theta}}\mathbf{G}\|^2$ can be represented as
    \begin{align}\label{beamforming_design_miso_5}
        \notag  \sum_{k=1}^{K}\|\mathbf{h}_k^{\mathrm{H}}\mathbf{\widetilde{\Theta}}\mathbf{G}\|^2=
        &\sum_{k=1}^{K}\|\mathbf{h}_{k}^\mathrm{H}\mathbf{J}_r\mathbf{\bar{\Theta}}\mathbf{J}_t\mathbf{G}\|^2\\ \notag = & \sum_{k=1}^{K}\|\mathbf{q}^{\mathrm{H}}\mathrm{diag}\{\mathbf{h}_{k}^\mathrm{H}\mathbf{J}_r\}\mathbf{J}_t\mathbf{G}\|^2\\
        \notag = & \sum_{k=1}^{K}\mathbf{q}^{\mathrm{H}}\mathbf{\Phi}_k\mathbf{\Phi}_k^{\mathrm{H}}\mathbf{q}\\
        = & \mathbf{q}^{\mathrm{H}}\left(\sum_{k=1}^{K}\mathbf{\Phi}_k\mathbf{\Phi}_k^{\mathrm{H}}\right)\mathbf{q},
    \end{align}
    where $\mathbf{\Phi}_k=\mathrm{diag}\{\mathbf{h}_{k}^\mathrm{H}\mathbf{J}_r\}\mathbf{J}_t\mathbf{G}$. Since the phase shift matrix $\mathbf{\widetilde{\Theta}}$ is determined by $\mathbf{q}$, $\mathbf{J}_t$ and $\mathbf{J}_r$, the problem of maximizing $\sum_{k=1}^{K}\|\mathbf{h}^{\mathrm{H}}\mathbf{\widetilde{\Theta}}\mathbf{G}\|^2$ is formulated as
    \begin{align}\label{beamforming_design_miso_6}
        \mathrm{(P3.b2)}\qquad&\max_{\mathbf{q},\mathbf{J}_t,\mathbf{J}_r}\mathbf{q}^{\mathrm{H}}\left(\sum_{k=1}^{K}\mathbf{\Phi}_k\mathbf{\Phi}_k^{\mathrm{H}}\right)\mathbf{q}\\
        \notag \mathrm{s.t.}&\quad |[\mathbf{q}]_i|=1,\quad i=1,2,\cdots,N.
    \end{align}
    In $\mathbf{\Phi}_k$, since the BS-RIS channel matrix $\mathbf{G}=[\mathbf{g}_1,\mathbf{g}_2,\dots,\mathbf{g}_M]$ has $M$ columns, and the RIS-users' channel matrix $\mathbf{h}=[\mathbf{h}_{1},\mathbf{h}_{2},\cdots,\mathbf{h}_{K}]^{\mathrm{H}}$ has $K$ rows, the permutation matrices $\mathbf{J}_t$ and $\mathbf{J}_r$ cannot be designed similarly to the SISO case by using a sorting method. Although the optimal permutation matrices $\mathbf{J}_t$ and $\mathbf{J}_r$ can be found by exhaustive search, we resort to the following sub-optimal method for reducing the search complexity. We introduce the notation of $\mathbf{g'}=\frac{1}{M}\sum_{m=1}^{M}|\mathbf{g}_m|$, and $\mathbf{h'}=\frac{1}{K}\sum_{k=1}^{K}|\mathbf{h}_{k}|$. Then, we choose the permutation matrix $\mathbf{J}_t$ which sorts the column vector $\mathbf{g'}$ in an ascending order, and the permutation matrix $\mathbf{J}_r$, which sorts the row vector $\mathbf{h'}^{\mathrm{H}}$ in an ascending order.

    After the permutation matrices $\mathbf{J}_t$ and $\mathbf{J}_r$ are determined, (P3.b2) becomes a non-convex quadratically constrained quadratic program (QCQP), and it can be solved by defining $\mathbf{Q}=\mathbf{q}\mathbf{q}^{\mathrm{H}}$, which needs to satisfy that $\mathbf{Q}\succeq0$ and $\mathrm{Rank}(\mathbf{Q}_\mathrm{opt})=1$. Since the rank-one constraint is non-convex, we relax this constraint \cite{wu2019intelligent} and (P3.b2) can be translated into a standard convex SDR problem as follows
    \begin{align}\label{beamforming_design_miso_7}
        \mathrm{(P3.b3)}\qquad&\max_{\mathbf{Q}}\mathrm{Tr}\left(\left(\sum_{k=1}^{K}\mathbf{\Phi}_k\mathbf{\Phi}_k^{\mathrm{H}}\right)\mathbf{Q}\right)\\
        \notag \mathrm{s.t.} &\quad \mathbf{Q}\succeq 0,\\
        \notag &\quad |[\mathbf{Q}]_{i,i}|=1,\quad i=1,2,\cdots,N.
    \end{align}
    The optimal solution, denoted as $\mathbf{Q}_\mathrm{opt}$, in (P3.b3) can be found by using CVX \cite{grant2008cvx}. If $\mathrm{Rank}(\mathbf{Q}_{\mathrm{opt}})>1$, the optimal solution of $\mathbf{q}$, denoted as $\mathbf{q}_{\mathrm{opt}}$, can be recovered from $\mathbf{Q}_{\mathrm{opt}}$ by eigenvalue decomposition as $\mathbf{q}_{\mathrm{opt}}=\sqrt{\nu_1}\mathbf{q}_1$, in which $\nu_1$ is the largest eigenvalue of the matrix $\mathbf{Q}_\mathrm{opt}$, and $\mathbf{q}_1$ is the corresponding eigenvector.

    After obtaining the permutation matrix $\mathbf{J}_t$, $\mathbf{J}_r$ and $\mathbf{q}_{\mathrm{opt}}$, the optimal phase shift matrix, denoted as $\mathbf{\widetilde{\Theta}}_\mathrm{opt}$, can be derived as $\mathbf{\widetilde{\Theta}}_\mathrm{opt}=\mathbf{J}_r\mathrm{diag}\{\mathbf{q}_\mathrm{opt}^{\mathrm{H}}\}\mathbf{J}_t$.
\item \textit{Stage II:} When the optimal phase shift matrix is obtained in the first stage, the equivalent channel can be represented as $\mathbf{H}_{\mathrm{equ}}=\mathbf{H}^{\mathrm{H}}\mathbf{\widetilde{\Theta}}_\mathrm{opt}\mathbf{G}$. Upon using the SVD method, the equivalent channel can be expressed as $\mathbf{H}_{\mathrm{equ}}=\mathbf{U}\mathbf{\Sigma}\mathbf{V}^{\mathrm{H}}$. The TBF matrix $\mathbf{W}$ is designed as $\mathbf{W}_{\mathrm{opt}}=\mathbf{V}^{(1:K)}$, where $\mathbf{V}^{(1:K)}$ represents the first $K$ columns of the right singular matrix $\mathbf{V}$. Afterwards, the power allocation matrix $\mathbf{\Lambda}$ is designed by the popular water-filling method \cite{tse2005fundamentals}, based on the optimized equivalent channel $\mathbf{H}_{\mathrm{equ}}$ and the optimized TBF matrix $\mathbf{W}_{\mathrm{opt}}$.
\end{itemize}

Finally, the achievable rate can be represented as
\begin{align}\label{beamforming_design_miso_8}
    R_{\mathrm{MIMO}}=\log_2\left(\mathbf{I}+\frac{P_t}{\sigma_n^2}\mathbf{H}_{\mathrm{equ}}\mathbf{C}\mathbf{H}_{\mathrm{equ}}^{\mathrm{H}}\right),
\end{align}
where $\mathbf{C}=\mathbf{W}_{\mathrm{opt}}\mathbf{\Lambda}\mathbf{W}_{\mathrm{opt}}^{\mathrm{H}}$ is the transmit covariance matrix.

\section{Theoretical Analysis}\label{Section_Theoretical_Analysis}
In this section, to highlight the channel gain enhancement in our proposed method compared with the conventional diagonal phase shift matrix method, we present the theoretical analysis of our proposed RIS architecture designed for the SISO systems. Firstly, we analyze the scaling law of our proposed RIS architecture relying on a non-diagonal phase matrix to derive the channel gain in Rician channels. Then, we discuss some special cases for different Rician factor values. Afterwards, we present the instantaneous SNR, as well as the outage probability and the average BER, of our proposed non-diagonal phase shift matrix based RIS systems in Rayleigh fading channels. Finally, the complexity comparison of the conventional RIS architecture and our proposed RIS architecture is presented.

\vspace{-4mm}
\subsection{Channel Gain Analysis}
To compare the fundamental limit of the conventional RIS architecture and of our proposed RIS architecture, we quantify the channel gain as a function of the number of RIS reflecting elements $N$.

In the SISO system, the BS-RIS channel vector and RIS-user channel vector are given by
\begin{align}\label{channel_gain_1}
    \mathbf{g}=\sqrt{\frac{\kappa_{\mathbf{g}}}{1+\kappa_{\mathbf{g}}}}\overline{\mathbf{g}}+\sqrt{\frac{1}{1+\kappa_{\mathbf{g}}}}\widetilde{\mathbf{g}},
\end{align}
\begin{align}\label{channel_gain_2}
    \mathbf{h}^\mathrm{H}=\sqrt{\frac{\kappa_{\mathbf{h}}}{1+\kappa_{\mathbf{h}}}}\overline{\mathbf{h}}_r^\mathrm{H}+\sqrt{\frac{1}{1+\kappa_{\mathbf{h}}}}\widetilde{\mathbf{h}}_r^\mathrm{H},
\end{align}
where $\kappa_{\mathbf{g}}$ and $\kappa_{\mathbf{h}}$ represent the Rician factors of BS-RIS path and RIS-user path, respectively. Since $a_i$ is the amplitudes of $\frac{1}{\sqrt{\varrho_t}}[\mathbf{g}]_i$, $a_i$ follows the Rice distribution with noncentrality parameter $\nu=\sqrt{\frac{\kappa_{\mathbf{g}}}{1+\kappa_{\mathbf{g}}}}$ and scale parameter $\sigma=\sqrt{\frac{1}{2\left(1+\kappa_{\mathbf{g}}\right)}}$, with the PDF and CDF as \cite{abdi2001estimation}
\begin{align}\label{channel_gain_6}
    f_{a_i}(x)=2\left(1+\kappa_\mathbf{g}\right)xe^{-\left(1+\kappa_\mathbf{g}\right)x^2-\kappa_\mathbf{g}}I_0\left(2\sqrt{\kappa_\mathbf{g}\left(1+\kappa_\mathbf{g}\right)}x\right),
\end{align}
\begin{align}\label{channel_gain_7}
    F_{a_i}(x)=1-Q_1\left(\sqrt{2\kappa_\mathbf{g}},\sqrt{2\left(1+\kappa_\mathbf{g}\right)}x\right),
\end{align}
where $I_0(\cdot)$ is the modified Bessel function of the first kind with order zero, and $Q_1(\cdot)$ is the Marcum Q-function \cite{abdi2001estimation}. Similarly, since $b_i$ is the amplitude of $\frac{1}{\sqrt{\varrho_r}}[\mathbf{h}^\mathrm{H}]_i$, $b_i$ follows Rice distribution with noncentrality parameter $\nu=\sqrt{\frac{\kappa_{\mathbf{h}}}{1+\kappa_{\mathbf{h}}}}$ and scale parameter $\sigma=\sqrt{\frac{1}{2\left(1+\kappa_{\mathbf{h}}\right)}}$, with the PDF and CDF as
\begin{align}\label{channel_gain_9}
    f_{b_i}(x)=2\left(1+\kappa_\mathbf{h}\right)xe^{-\left(1+\kappa_\mathbf{h}\right)x^2-\kappa_\mathbf{h}}I_0\left(2\sqrt{\kappa_\mathbf{h}\left(1+\kappa_\mathbf{h}\right)}x\right),
\end{align}
\begin{align}\label{channel_gain_10}
    F_{b_i}(x)=1-Q_1\left(\sqrt{2\kappa_\mathbf{h}},\sqrt{2\left(1+\kappa_\mathbf{h}\right)}x\right).
\end{align}
Since in a Rice distribution $\mathrm{Rice}\left(\nu,\sigma\right)$, the first moment and second moment are $\sigma\sqrt{\pi/2}L_{1/2}\left(-\nu^2/2\sigma^2\right)$ and $\nu^2+2\sigma^2$ respectively \cite{abdi2001estimation}, in which $L_{1/2}\left(\cdot\right)$ is the Laguerre polynomial, we can get that \cite{abdi2001estimation}
\begin{align}\label{channel_gain_11}
    \mathbb{E}\left(a_i\right)=\sqrt{\frac{\pi}{4\left(1+\kappa_{\mathbf{g}}\right)}}L_{\frac{1}{2}}\left(-\kappa_\mathbf{g}\right),
\end{align}
\begin{align}\label{channel_gain_12}
    \mathbb{E}\left(b_i\right)=\sqrt{\frac{\pi}{4\left(1+\kappa_{\mathbf{h}}\right)}}L_{\frac{1}{2}}\left(-\kappa_{\mathbf{h}}\right),
\end{align}
\begin{align}\label{channel_gain_13}
    \mathbb{E}\left(a_i^2\right)=\mathbb{E}\left(b_i^2\right)=1,
\end{align}

\subsubsection{Average channel gain in the conventional RIS architecture}
In SISO systems, the average channel gain of the conventional RIS architecture relying on the diagonal phase shift matrix, denoted as $\mathfrak{g}_{\mathrm{diag}}$, when considering Rician fading channel, is given by
\begin{align}\label{Rician_scaling_law_conven_1}
    \notag \mathfrak{g}_{\mathrm{diag}}&=\mathbb{E}\left(|\mathbf{h}^\mathrm{H}\mathbf{\bar{\Theta}} \mathbf{g}|^2\right)\\
    \notag &=\mathbb{E}\left(\left(\sum_{i=1}^{N}|[\mathbf{g}]_i||[\mathbf{h}^\mathrm{H}]_i|\right)^2\right)\\
    \notag &=\varrho_t\varrho_r\mathbb{E}\left(\left(\sum_{i=1}^{N}a_ib_i\right)^2\right)\\
    \notag&=\varrho_t\varrho_r\Bigg(\sum_{i=1}^{N}\mathbb{E}\left(a_i^2\right)\mathbb{E}\left(b_i^2\right)\\
    &\quad+2\sum_{i=1}^{N-1}\sum_{j=i+1}^{N}\mathbb{E}\left(a_i\right)\mathbb{E}\left(b_i\right)\mathbb{E}\left(a_j\right)\mathbb{E}\left(b_j\right)\Bigg).
\end{align}
According to (\ref{channel_gain_11}), (\ref{channel_gain_12}), (\ref{channel_gain_13}) and (\ref{Rician_scaling_law_conven_1}), we can get
\begin{align}\label{Rician_scaling_law_conven_2}
    \mathfrak{g}_{\mathrm{diag}}\!=\!\varrho_t\varrho_r\!\left(N\!+\!\frac{\pi^2}{16}N\!(N\!-\!1)\frac{L_{\frac{1}{2}}^2\left(-\kappa_\mathbf{g}\right)L_{\frac{1}{2}}^2\left(-\kappa_{\mathbf{h}}\right)}{\left(\kappa_\mathbf{g}+1\right)\left(\kappa_{\mathbf{h}}+1\right)}\right).
\end{align}

\subsubsection{Average channel gain analysis in the proposed RIS architecture}
Since $a_{(1)},a_{(2)},\cdots,a_{(N)}$ and $b_{(1)},b_{(2)},\cdots,b_{(N)}$ represent the sequences of $a_1,a_2,\cdots,a_N$ and $b_1,b_2,\cdots,b_N$ sorted in an ascending order, the average channel gain of the proposed RIS architecture having a non-diagonal phase shift matrix, denoted as $\mathfrak{g}_\mathrm{nond}$, can be expressed as
\begin{align}\label{Rician_scaling_law_sort_1}
    \notag &\mathfrak{g}_\mathrm{nond}= \mathbb{E}\left(|\mathbf{h}^{\mathrm{H}}\mathbf{J}_r\mathbf{\bar{\Theta}}\mathbf{J}_t\mathbf{g}|^2\right)\\
    \notag &\qquad\ =\varrho_t\varrho_r \mathbb{E}\left(\left(\sum_{i=1}^{N}a_{(i)}b_{(i)}\right)^2\right)\\
    \notag &\qquad\ = \varrho_t\varrho_r\Bigg(\sum_{i=1}^{N}\mathbb{E}\left(a_{(i)}^2\right)\mathbb{E}\left(b_{(i)}^2\right)\\
    &\quad+2\sum_{i=1}^{N-1}\sum_{j=i+1}^{N}\mathbb{E}\left(a_{(i)}\right)\mathbb{E}\left(b_{(i)}\right)\mathbb{E}\left(a_{(j)}\right)\mathbb{E}\left(b_{(j)}\right)\Bigg).
\end{align}
According to the order statistic theory \cite{renyi1953theory}, we can get the PDF of $a_{(i)}$ and $b_{(i)}$ as
\begin{align}\label{Rician_scaling_law_sort_2}
    f_{a_{(i)}}(x)=\mathrm{C}_{i,N}\left[F_{a_i}(x)\right]^{i-1}\left[1-F_{a_i}(x)\right]^{n-i}f_{a_i}(x),
\end{align}
\begin{align}\label{Rician_scaling_law_sort_3}
    f_{b_{(i)}}(x)=\mathrm{C}_{i,N}\left[F_{b_i}(x)\right]^{i-1}\left[1-F_{b_i}(x)\right]^{n-i}f_{b_i}(x),
\end{align}
where $f_{a_i}(x)$, $F_{a_i}(x)$, $f_{b_i}(x)$ and $F_{b_i}(x)$ are given in (\ref{channel_gain_6}), (\ref{channel_gain_7}), (\ref{channel_gain_9}) and (\ref{channel_gain_10}), respectively, and $\mathrm{C}_{i,N}=\frac{N!}{(i-1)!(N-i)!}$. Therefore, the first moment of $a_{(i)}$ is given by
\begin{align}\label{Rician_scaling_law_sort_4}
    \notag\mathbb{E}\left(a_{(i)}\right)&=\int_0^{\infty}xf_{a_{(i)}}(x)\mathrm{d}x\\
    &=\int_0^{\infty}x\mathrm{C}_{i,N}\left[F_{a_i}(x)\right]^{i-1}\left[1-F_{a_i}(x)\right]^{n-i}f_{a_i}(x)\mathrm{d}x.
\end{align}
Substituting (\ref{channel_gain_6}) and (\ref{channel_gain_7}) into (\ref{Rician_scaling_law_sort_4}), we can get
\begin{align}\label{Rician_scaling_law_sort_5}
    \notag&\mathbb{E}\left(a_{(i)}\right)=\frac{\mathrm{C}_{i,N}e^{-\kappa_{\mathbf{g}}}}{\sqrt{2\left(1+\kappa_{\mathbf{g}}\right)}}\int_0^{\infty}x^2e^{-\frac{x^2}{2}}I_0\left(\sqrt{2\kappa_\mathbf{g}}x\right)\times\\
    &\quad\left[1-Q_1\left(\sqrt{2\kappa_{\mathbf{g}}},x\right)\right]^{i-1}\left[Q_1\left(\sqrt{2\kappa_{\mathbf{g}}},x\right)\right]^{N-i}\mathrm{d}x.
\end{align}
Similarly, we can get the second moment of $a_{(i)}$, the first moment of $b_{(i)}$, and the second moment of $b_{(i)}$ as
\begin{align}\label{Rician_scaling_law_sort_6}
    \notag&\mathbb{E}\left(a_{(i)}^2\right)=\int_0^{\infty}x^2f_{a_{(i)}}(x)\mathrm{d}x\\
    \notag&\qquad\quad\ =\frac{\mathrm{C}_{i,N}e^{-\kappa_{\mathbf{g}}}}{2\left(1+\kappa_{\mathbf{g}}\right)}\int_0^{\infty}x^3e^{-\frac{x^2}{2}}I_0\left(\sqrt{2\kappa_\mathbf{g}}x\right)\times\\
    &\quad\left[1-Q_1\left(\sqrt{2\kappa_{\mathbf{g}}},x\right)\right]^{i-1}\left[Q_1\left(\sqrt{2\kappa_{\mathbf{g}}},x\right)\right]^{N-i}\mathrm{d}x,
\end{align}
\begin{align}\label{Rician_scaling_law_sort_7}
    \notag&\mathbb{E}\left(b_{(i)}\right)=\int_0^{\infty}xf_{b_{(i)}}(x)\mathrm{d}x\\
    \notag&\qquad\quad\ =\frac{\mathrm{C}_{i,N}e^{-\kappa_{\mathbf{h}}}}{\sqrt{2\left(1+\kappa_{\mathbf{h}}\right)}}\int_0^{\infty}x^2e^{-\frac{x^2}{2}}I_0\left(\sqrt{2\kappa_\mathbf{h}}x\right)\times\\
    &\quad\left[1-Q_1\left(\sqrt{2\kappa_{\mathbf{h}}},x\right)\right]^{i-1}\left[Q_1\left(\sqrt{2\kappa_{\mathbf{h}}},x\right)\right]^{N-i}\mathrm{d}x,
\end{align}
\begin{align}\label{Rician_scaling_law_sort_8}
    \notag&\mathbb{E}\left(b_{(i)}^2\right)=\int_0^{\infty}x^2f_{b_{(i)}}(x)\mathrm{d}x\\
    \notag&\qquad\quad\ =\frac{\mathrm{C}_{i,N}e^{-\kappa_{\mathbf{h}}}}{2\left(1+\kappa_{\mathbf{h}}\right)}\int_0^{\infty}x^3e^{-\frac{x^2}{2}}I_0\left(\sqrt{2\kappa_\mathbf{h}}x\right)\times\\
    &\quad\left[1-Q_1\left(\sqrt{2\kappa_{\mathbf{h}}},x\right)\right]^{i-1}\left[Q_1\left(\sqrt{2\kappa_{\mathbf{h}}},x\right)\right]^{N-i}\mathrm{d}x.
\end{align}
Substituting (\ref{Rician_scaling_law_sort_5}), (\ref{Rician_scaling_law_sort_6}), (\ref{Rician_scaling_law_sort_7}) and (\ref{Rician_scaling_law_sort_8}) into (\ref{Rician_scaling_law_sort_1}), we can get the average channel gain of the proposed RIS
architecture $\mathfrak{g}_\mathrm{nond}$.

\vspace{-5mm}
\subsection{Effect of the Value of the Rician Factor on the Channel Gain}
To get deep insights on the effects of the Rician factors $\kappa_{\mathbf{g}}$ and $\kappa_{\mathbf{h}}$ on the channel gain, we investigate the following cases.

\subsubsection*{Case I, $\kappa_{\mathbf{g}}\rightarrow\infty$ (or $\kappa_{\mathbf{h}}\rightarrow\infty$)}
In this case, the BS-RIS channel (or RIS-user channel) is fully dominated by the LoS path, the conventional RIS architecture and our proposed RIS architecture get the same average channel gain as
\begin{align}\label{K_Effect_0}
    \mathfrak{g}_\mathrm{diag}=\mathfrak{g}_\mathrm{nond}.
\end{align}
\begin{IEEEproof}
See Appendix \ref{Appendix_1}.
\end{IEEEproof}
Furthermore, when $\kappa_{\mathbf{g}}\rightarrow\infty$ and $\kappa_{\mathbf{h}}\rightarrow\infty$ simultaneously, both the conventional RIS architecture and our proposed RIS architecture gets the channel gain upper bound as
\begin{align}\label{K_Effect_2}
    \mathfrak{g}_\mathrm{diag}=\mathfrak{g}_\mathrm{nond}=\varrho_t\varrho_rN^2.
\end{align}
This can be easily observed since $a_i=1$ and $b_i=1$ for all $i=1,2,\cdots,N$ when $\kappa_{\mathbf{g}}\rightarrow\infty$ and $\kappa_{\mathbf{h}}\rightarrow\infty$.

\subsubsection*{Case II, $\kappa_{\mathbf{g}}=0$ and $\kappa_{\mathbf{h}}=0$}
In this case, the BS-RIS channel and RIS-user channel experience Rayleigh fading. The average channel gain of the conventional RIS architecture $\mathfrak{g}_\mathrm{diag}$ and the proposed non-diagonal RIS architecture $\mathfrak{g}_\mathrm{nond}$ are derived as follows.

\begin{itemize}
  \item \textit{Average channel gain in the conventional RIS architecture:} When $\kappa_{\mathbf{g}}=0$ and $\kappa_{\mathbf{h}}=0$, $a_i$ and $b_i$ both follow the Rayleigh distribution associated with the scaling parameter $\sigma=\frac{\sqrt{2}}{2}$, and $a_i^2$ and $b_i^2$ obey the exponential distribution having the rate parameter of $\lambda=1$. Therefore,
      \begin{align}\label{scaling_law_conven_0}
        \mathbb{E}\left(a_i\right)=\mathbb{E}\left(b_i\right)=\frac{\sqrt{\pi}}{2},
      \end{align}
      \begin{align}\label{scaling_law_conven_1}
        \mathbb{E}\left(a_i^2\right)=\mathbb{E}\left(b_i^2\right)=1.
      \end{align}
      Substituting (\ref{scaling_law_conven_0}) and (\ref{scaling_law_conven_1}) into (\ref{Rician_scaling_law_conven_1}), we can get
      \begin{align}\label{scaling_law_conven_2}
        \mathfrak{g}_{\mathrm{diag}}=\varrho_t\varrho_r\left(N+N(N-1)\frac{\pi^2}{16}\right).
      \end{align}
  \item \textit{Average channel gain in the proposed RIS architecture:} Since $a_i$ and $b_i$ both follow Rayleigh distribution associated with the scaling parameter $\sigma=\frac{\sqrt{2}}{2}$, $a_{(i)}$ and $b_{(i)}$ are also identically distributed. Therefore, (\ref{Rician_scaling_law_sort_1}) can be simplified as
      \begin{align}\label{scaling_law_proposed_1}
        \notag&\mathfrak{g}_\mathrm{nond}=\varrho_t\varrho_r\Bigg(\sum_{i=1}^{N}\left(\mathbb{E}\left(a_{(i)}^2\right)\right)^2\\
        &\qquad+2\sum_{i=1}^{N-1}\sum_{j=i+1}^{N}\left(\mathbb{E}\left(a_{(i)}\right)\right)^2\left(\mathbb{E}\left(a_{(j)}\right)\right)^2\Bigg).
      \end{align}

      Note that $a_1^2,a_2^2,\cdots,a_N^2$ independently follow the exponential distribution having the rate parameter $\lambda=1$, and given the lemma in \cite{renyi1953theory} that
      \begin{align}\label{scaling_law_proposed_2}
        a_{(i)}^2=\sum_{k=1}^i \frac{Z_k}{N-k+1},\quad i=1,2,\cdots,N,
      \end{align}
      where the random variable $Z_k$ follows the exponential distribution with the rate parameter $\lambda=1$, we arrive at:
      \begin{align}\label{scaling_law_proposed_3}
        \sum_{i=1}^{N}\left(\mathbb{E}\left(a_{(i)}^2\right)\right)^2=\sum_{i=1}^N\left(\sum_{k=1}^i \frac{1}{N-k+1}\right)^2.
      \end{align}

      To derive $\sum_{i=1}^{N-1}\sum_{j=i+1}^{N}\left(\mathbb{E}\left(a_{(i)}\right)\right)^2\left(\mathbb{E}\left(a_{(j)}\right)\right)^2$, firstly, we present the following \textit{Theorem} \ref{theorem_1}.

      \begin{theorem}\label{theorem_1}
        The PDF of $a_{(i)}$ is given by a linear combination of the PDF of $i$ Rayleigh distributions having the scaling parameter $\sigma=\frac{1}{\sqrt{2(N-i+k+1)}}$. Specifically, the PDF of $a_{(i)}$ can be written as
        \begin{align}\label{scaling_law_proposed_4}
            \notag f_{a_{(i)}}(x)&=\sum_{k=0}^{i-1}\binom{N}{i-k-1}\binom{N-i+k}{k}(-1)^k\\
            &f_{\mathrm{X}}\left(x;\sigma=\frac{1}{\sqrt{2(N-i+k+1)}}\right),
        \end{align}
        where $f_{\mathrm{X}}(x;\sigma)$ is the PDF of a random variable $X$ following the Rayleigh distribution having the scaling parameter $\sigma$.
      \end{theorem}

      \begin{IEEEproof}
        See Appendix \ref{Appendix_2}.
      \end{IEEEproof}

      Since the mean of $f_X(x;\sigma)$ equals $\sigma\sqrt{\frac{\pi}{2}}$, then
      \begin{align}\label{scaling_law_proposed_8}
        \notag &\mathbb{E}\left(f_{\mathrm{X}}(x;\sigma=\frac{1}{\sqrt{2(N-i+k+1)}})\right)\\
        =&\frac{\sqrt{\pi}}{2}\frac{1}{\sqrt{N-i+k+1}}.
      \end{align}
      Upon combining (\ref{scaling_law_proposed_4}) and (\ref{scaling_law_proposed_8}), we arrive at:
      \begin{align}\label{scaling_law_proposed_9}
        \notag &\mathbb{E}\left[f_{a_{(i)}}(x)\right]=\frac{\sqrt{\pi}}{2}\sum_{k=0}^{i-1}\binom{N}{i-k-1}\binom{N-i+k}{k}\\
        &\qquad (-1)^k\frac{1}{\sqrt{N-i+k+1}}.
      \end{align}

      Therefore, $\sum_{i=1}^{N-1}\sum_{j=i+1}^{N}\left[\mathbb{E}\left(a_{(i)}\right)\right]^2\left[\mathbb{E}\left(a_{(j)}\right)\right]^2$ is obtained as
      \begin{align}\label{scaling_law_proposed_10}
            \notag&\sum_{i=1}^{N-1}\sum_{j=i+1}^{N}\left(\mathbb{E}\left(a_{(i)}\right)\right)^2\left(\mathbb{E}\left(a_{(j)}\right)\right)^2=\frac{\pi^2}{16}\sum_{i=1}^{N-1}\sum_{j=i+1}^{N}\\
            \notag&\left[\sum_{k=0}^{i-1}\!(-1)^k\!\binom{N}{i\!-\!k\!-\!1}\!\binom{N\!-\!i\!+\!k}{k}\!\frac{1}{\sqrt{N\!-\!i\!+\!k\!+\!1}}\right]^2\\
            &\left[\sum_{k=0}^{j-1}(-1)^k\!\binom{N}{j\!-\!k\!-\!1}\!\binom{N\!-\!j\!+\!k}{k}\!\frac{1}{\sqrt{N\!-\!j\!+\!k\!+\!1}}\right]^2.
        \end{align}

      According to (\ref{scaling_law_proposed_1}), (\ref{scaling_law_proposed_3}) and (\ref{scaling_law_proposed_10}), we get the theoretical result of the channel gain $\mathfrak{g}_{\mathrm{nond}}$ as
      \begin{align}\label{scaling_law_proposed_11}
            \notag &\mathfrak{g}_{\mathrm{nond}}=\varrho_t\varrho_r\sum_{i=1}^N\left(\sum_{k=1}^i \frac{1}{N\!-\!k\!+\!1}\right)^2\!+\!\varrho_t\varrho_r\frac{\pi^2}{8}\sum_{i=1}^{N-1}\sum_{j=i+1}^{N}\\
            \notag&\left[\sum_{k=0}^{i-1}\binom{N}{i\!-\!k\!-\!1}\binom{N\!-\!i\!+\!k}{k}(-1)^k\frac{1}{\sqrt{N\!-\!i\!+\!k\!+\!1}}\right]^2\\
            &\left[\sum_{k=0}^{j-1}\binom{N}{j\!-\!k\!-\!1}\binom{N\!-\!j\!+\!k}{k}(-1)^k\frac{1}{\sqrt{N\!-\!j\!+\!k\!+\!1}}\right]^2.
      \end{align}

      Upon comparing (\ref{Rician_scaling_law_conven_1}) to (\ref{Rician_scaling_law_sort_1}), we can find that the channel gain of the conventional RIS architecture is proportional to the expectation of the square of $a_1b_1+a_2b_2+\cdots+a_Nb_N$, while the channel gain of our proposed RIS architecture is proportional to the expectation of the square of $a_{(1)}b_{(1)}+a_{(2)}b_{(2)}+\cdots+a_{(N)}b_{(N)}$. Therefore, in conventional RIS architectures the channel gain is proportional to the equal gain combination of the channel parameters, while in our proposed RIS architecture the channel gain is proportional to the maximum ratio combination of the channel parameters, when the number of RIS elements $N$ is large.

      We define the \textit{normalized channel gain} $\mathfrak{\widehat{g}}$ as the average channel gain $\mathfrak{g}$ normalized by the upper bound of $\varrho_t\varrho_rN^2$ in LoS channels, i.e. $\mathfrak{\widehat{g}}=\frac{\mathfrak{g}}{\varrho_t\varrho_rN^2}$. In Fig. \ref{Figure_5}, we present the normalized power gain of our proposed RIS architecture, compared to that of the fully-connected/group-connected RIS architecture of \cite{shen2020modeling} in Rayleigh fading channels, i.e. $\kappa_\mathbf{g}\!=\!\kappa_\mathbf{h}\!=\!0$, where $G$ represents the group size of a group-connected RIS architecture. Fig. \ref{Figure_5} shows that when the number of RIS elements is small, the power gain of our proposed RIS architecture is worse than that of the fully connected RIS architecture of \cite{shen2020modeling}. However, the performance of our proposed RIS architecture becomes better than that of the group-connected RIS architecture and approaches that of the fully-connected RIS architecture upon increasing the number of RIS reflecting elements $N$. However, regardless of the number of RIS elements $N$, our RIS architecture outperforms the conventional RIS architecture. This gain is due to the fact that in our RIS architecture, the BS-RIS channel vector $\mathbf{g}$ and the RIS-user channel vector $\mathbf{h}^{\mathrm{H}}$ are combined based on the MRC criterion, when the number of elements $N$ is large.

    \begin{figure}[!t]
    \vspace{-0mm}
    \setstretch{0.8}
    \setlength{\abovecaptionskip}{-3pt}
    \setlength{\belowcaptionskip}{-3pt}
        \centering
        \includegraphics[width=3.95in]{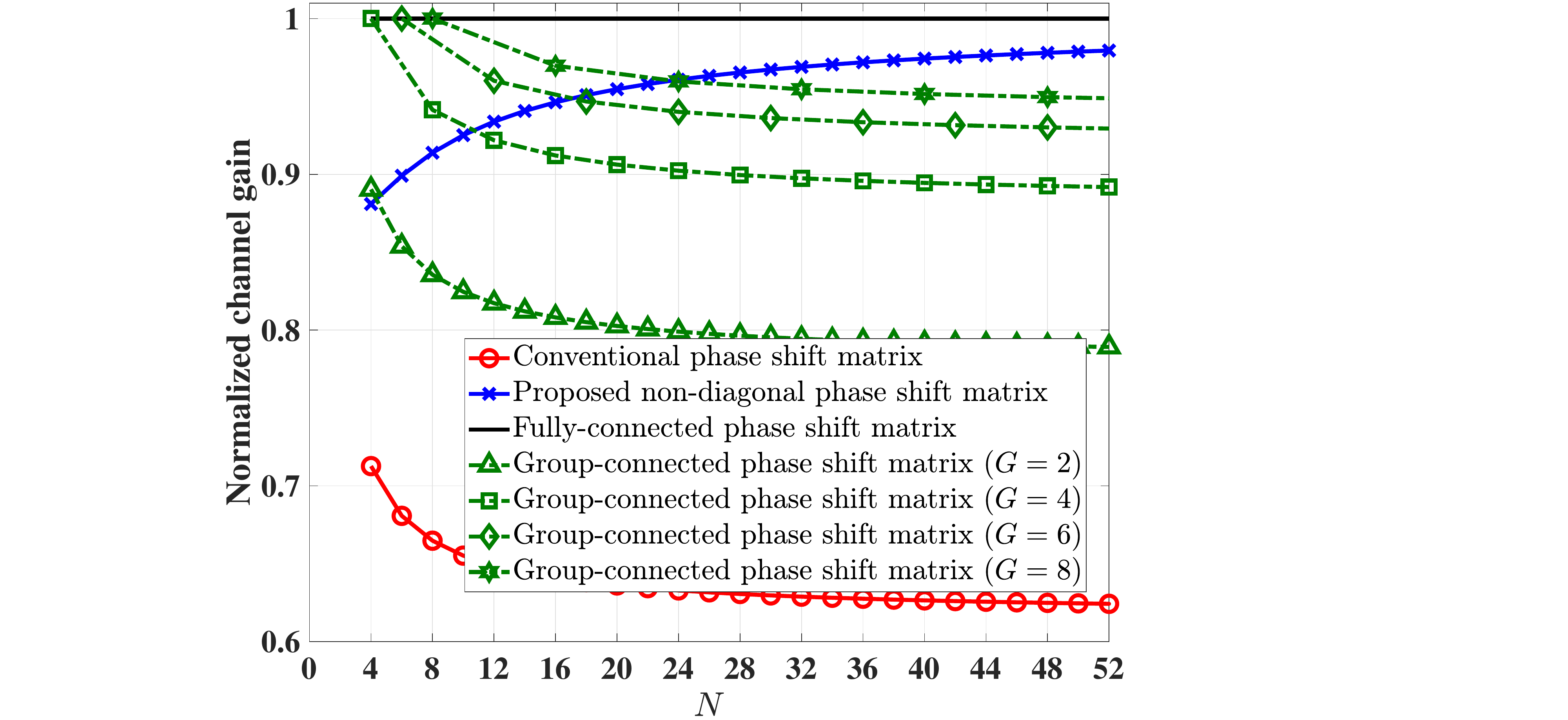}
        \caption{Comparison of normalized channel gain versus the number of RIS elements $N$ for the different RIS architectures.}\label{Figure_5}
    \vspace{-3mm}
    \end{figure}
\end{itemize}

\vspace{-5mm}
\subsection{Channel Gain Analysis when $N\rightarrow\infty$}
In the conventional RIS architecture, when $N\rightarrow\infty$, according to (\ref{Rician_scaling_law_conven_2}), we can get the average channel gain as
\begin{align}\label{N_infty_1}
    \mathfrak{g}_{\mathrm{diag}}=\frac{\pi^2}{16}\varrho_t\varrho_rN^2\frac{L_{\frac{1}{2}}^2\left(-\kappa_\mathbf{g}\right)L_{\frac{1}{2}}^2\left(-\kappa_{\mathbf{h}}\right)}{\left(\kappa_\mathbf{g}+1\right)\left(\kappa_{\mathbf{h}}+1\right)}.
\end{align}

In our proposed RIS architecture, when $N\rightarrow\infty$, the average channel gain is given by
\begin{align}\label{N_infty_2}
    \mathfrak{g}_\mathrm{nond}=\varrho_t\varrho_rN^2,
\end{align}
which is equivalent to the channel gain performance of the fully-connected RIS architecture of \cite{shen2020modeling}.
\begin{IEEEproof}
See Appendix \ref{Appendix_3}.
\end{IEEEproof}

Fig. \ref{Figure_6} compares the normalized channel gain versus BS-RIS Rician factors $\kappa_\mathbf{g}$ and RIS-user Rician factors $\kappa_\mathbf{h}$ for the conventional RIS architecture and our proposed RIS architectures. It shows that when $N\rightarrow\infty$, the normalized channel gain of the conventional RIS architecture degrades with the decrease of the Rician factors, while that of our proposed RIS architecture remains at 1 in all Rician factor ranges, which means that our proposed RIS architecture is more robust over a wider range of propagation conditions and especially shows advantages in NLoS-dominated channel environments.

\begin{figure}[!t]
\vspace{-0mm}
\setstretch{1}
\setlength{\abovecaptionskip}{-3pt}
\setlength{\belowcaptionskip}{-3pt}
    \centering
    \includegraphics[width=3.75in]{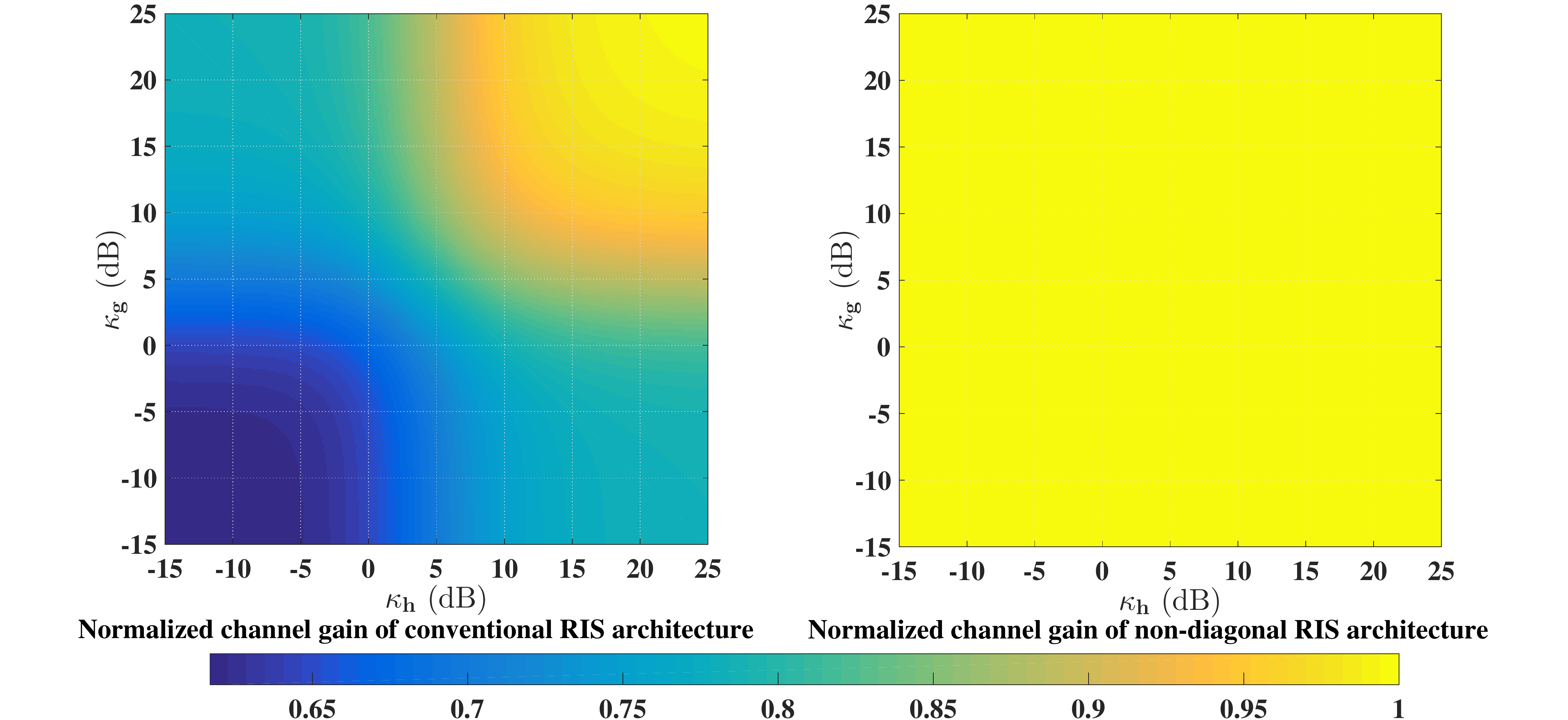}
    \caption{Comparison of normalized channel gain versus BS-RIS Rician factors $\kappa_\mathbf{g}$ and RIS-user Rician factors $\kappa_\mathbf{h}$ for the conventional RIS architecture and our proposed RIS architectures.}\label{Figure_6}
\vspace{-5mm}
\end{figure}

\vspace{-5mm}
\subsection{Outage Probability and BER Performance Analysis}
Since the above analysis demonstrates that our proposed RIS architecture shows advantages in NLoS-dominated channel environments, in this section we derive the distribution of the received SNR of our proposed RIS architecture in Rayleigh fading channels, then the outage probability and BER performance are analyzed.

According to (\ref{N_infty_3}), we can show that the upper bound of the instantaneous SNR at the receiver side of our proposed RIS architecture can be expressed as
\begin{align}\label{Receiver_SNR_1}
    \Omega=\rho\left(\sum_{i=1}^{N}a_i^2\right)\left(\sum_{i=1}^{N}b_i^2\right),
\end{align}
where we have $\rho=\frac{P_t}{\sigma_n^2}\varrho_{t}\varrho_{r}$. Since $a_i^2$ and $b_i^2$ both obey the exponential distribution having the rate parameter of $\lambda=1$, then both $Y_a=\sum_{i=1}^{N}a_i^2$ and $Y_b=\sum_{i=1}^{N}b_i^2$ follow the gamma distribution associated with the shape parameter $N$ and the scale parameter 1. Let us introduce $Z=Y_aY_b$, with the following PDF \cite{withers2013product}
\begin{align}\label{Receiver_SNR_2}
    f_Z(z)=\frac{2}{\Gamma^2(N)}z^{N-1}K_0(2\sqrt{z}),
\end{align}
where $\Gamma(\cdot)$ is the gamma function, and $K_0(\cdot)$ is the modified Bessel function of the second kind. Therefore, the CDF of the received SNR $\Omega$ is given by
\begin{align}\label{Receiver_SNR_5}
    F_\Omega(\omega)=\int_0^{\frac{\omega}{\rho}}\frac{2}{\Gamma^2(N)}t^{N-1}K_0\left(2\sqrt{t}\right)\mathrm{d}t.
\end{align}

\subsubsection{Outage probability}
If the SNR threshold is $\omega_{\mathrm{th}}$, then the outage probability, denoted as $P_{\mathrm{out}}(\omega_{\mathrm{th}})$, can be calculated as \cite{tse2005fundamentals}
\begin{align}\label{Receiver_SNR_6}
    P_{\mathrm{out}}(\omega_{\mathrm{th}})\!=\!\mathrm{Pr}(\Omega\leq \omega_{\mathrm{\mathrm{th}}})\!=\!\int_0^{\frac{\omega_\mathrm{th}}{\rho}}\!\!\frac{2}{\Gamma^2(N)}t^{N-1}K_0\left(2\sqrt{t}\right)\mathrm{d}t.
\end{align}

\subsubsection{Average BER}
Based on the CDF of the received SNR in (\ref{Receiver_SNR_5}), the average BER is given by \cite{ansari2011new}
\begin{align}\label{Receiver_SNR_7}
    \notag P_e&=\frac{q^p}{2\Gamma(p)}\int_{0}^{\infty}e^{-q\omega}\omega^{p-1}F_\Omega(\omega)\mathrm{d}\omega\\ &=\frac{q^p}{\Gamma(p)\Gamma^2(N)}\int_{0}^{\infty}e^{-q\omega}\omega^{p-1}\mathrm{d}\omega\int_0^{\frac{\omega}{\rho}}t^{N-1}K_0\left(2\sqrt{t}\right)\mathrm{d}t,
\end{align}
where the parameters $p$ and $q$ are different for different modulation schemes. For example, we have $p=\frac{1}{2}$ and $q=1$ for binary phase shift keying (BPSK) modulation \cite{ansari2011new}.

\vspace{-3mm}
\subsection{Complexity Analysis}
In this section, we analyze the complexity of the RIS architecture in terms of the number of configurable impedances required and the BS-RIS control link load.

\begin{figure*}[htb]
\vspace{-0mm}
\setstretch{0.8}
\setlength{\abovecaptionskip}{-0pt}
\setlength{\belowcaptionskip}{-0pt}
    \begin{minipage}[t]{0.495\linewidth}
        \centering
        \includegraphics[width=1.25\textwidth]{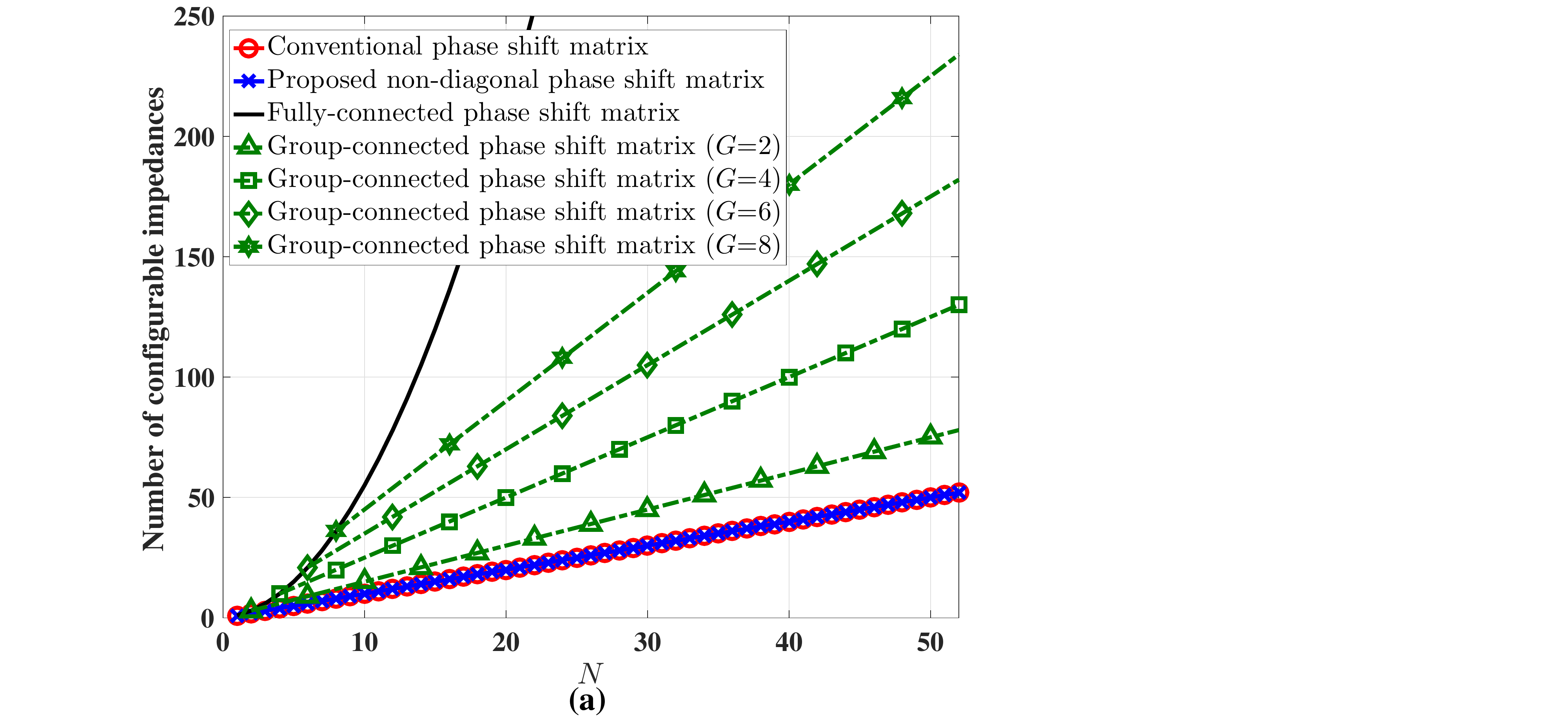}
    \end{minipage}
    \begin{minipage}[t]{0.495\linewidth}
        \centering
        \includegraphics[width=1.25\textwidth]{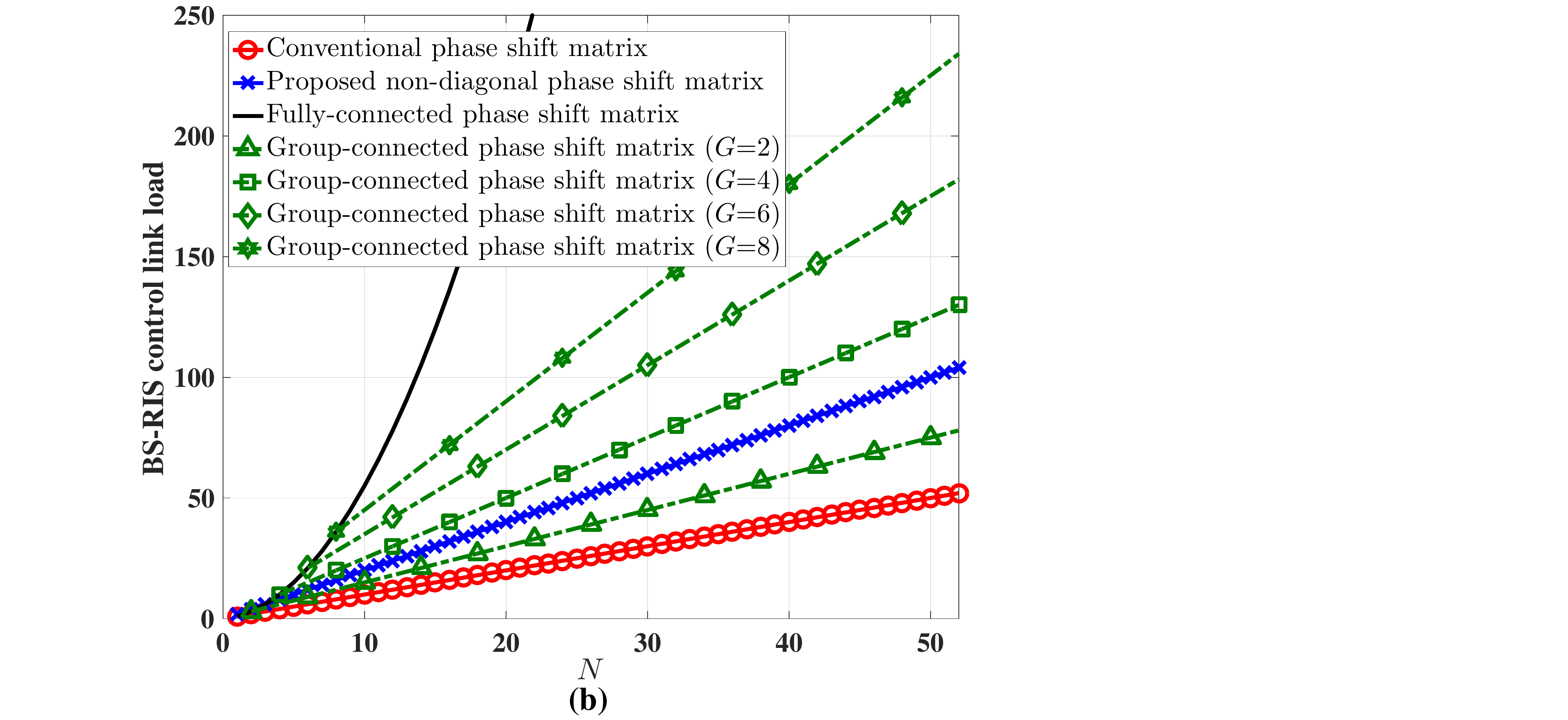}
    \end{minipage}
    \caption{(a) Comparison of the number of configurable impedances versus the number of RIS elements $N$ for the different RIS architectures. (b) Comparison of the BS-RIS control link load versus the number of RIS elements $N$ for the different RIS architectures.}\label{Figure_7}
\vspace{-3mm}
\end{figure*}

Firstly, the conventional RIS architecture requires $N$ configurable impedances, while for the fully-connected RIS architecture we need $\frac{N(N+1)}{2}$ and for the group-connected RIS architecture we need $\frac{N(G+1)}{2}$. By contrast, our architecture only requires $N$ configurable impedances, which is the same as that in the conventional RIS architecture. Fig. \ref{Figure_7} (a) compares the number of configurable impedances required for the conventional RIS structure, the fully-connected/group-connected RIS structure and our RIS structure, which shows that the number of configurable impedances required by our proposed RIS architecture is lower than that of the fully-connected RIS architecture and of the group-connected RIS architecture.

Secondly, in terms of the BS-RIS control link load, the number of information values transmitted on the BS-RIS control link is $N$ for the conventional RIS architecture, since only the diagonal values of the phase shift matrix are optimized, while for the fully-connected RIS architecture it is $\frac{N(N+1)}{2}$ and for the group-connected RIS architecture it is $\frac{N(G+1)}{2}$. In our proposed non-diagonal phase shift matrix scheme, the number of information values transmitted in BS-RIS control link is $2N$, i.e., $N$ optimized non-zeros values and their positions in the non-diagonal phase shift matrix. Fig. \ref{Figure_7} (b) compares the BS-RIS control link load for the conventional RIS structure, the fully-connected/group-connected RIS structure and our RIS structure, which shows that the BS-RIS control link load of our RIS architecture is lower than that of the fully-connected and of the group-connected RIS architecture for $G\geq4$. However, it has been presented in Fig. \ref{Figure_5} that the performance of our proposed RIS structure is better than that of the group-connected method and approaches that of the fully-connected RIS structure, as the number of RIS elements increases.

\section{Performance results and analysis}\label{Section_Performance_results_and_analysis}
In this section, firstly the outage probability and average BER of our proposed RIS architecture are presented. Then, we characterize the achievable rate of our proposed RIS architecture having non-diagonal phase shift matrices.

\begin{figure*}[htb]
\vspace{-0mm}
\setstretch{0.8}
\setlength{\abovecaptionskip}{-3pt}
\setlength{\belowcaptionskip}{-3pt}
    \begin{minipage}[t]{0.495\linewidth}
        \centering
        \includegraphics[width=1.25\textwidth]{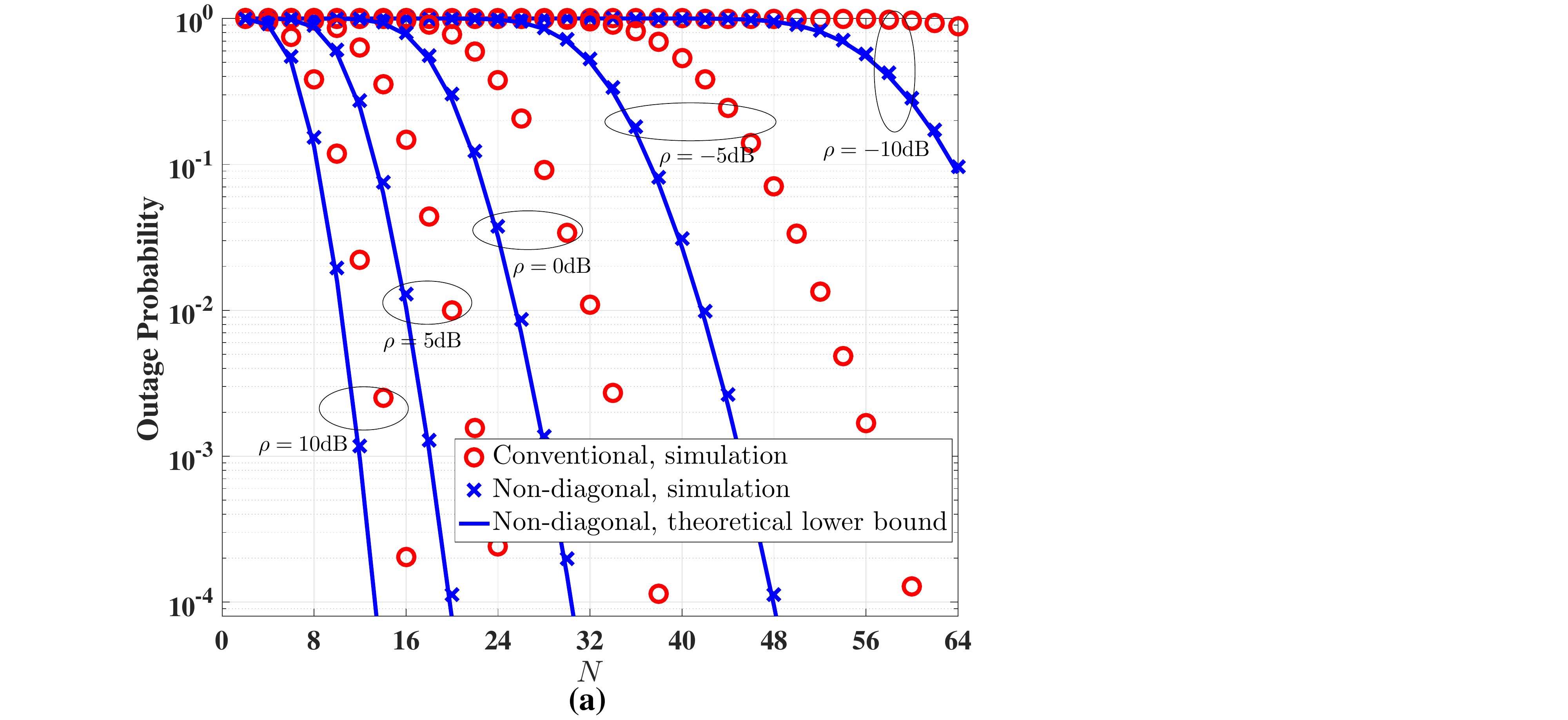}
    \end{minipage}
    \begin{minipage}[t]{0.495\linewidth}
        \centering
        \includegraphics[width=1.25\textwidth]{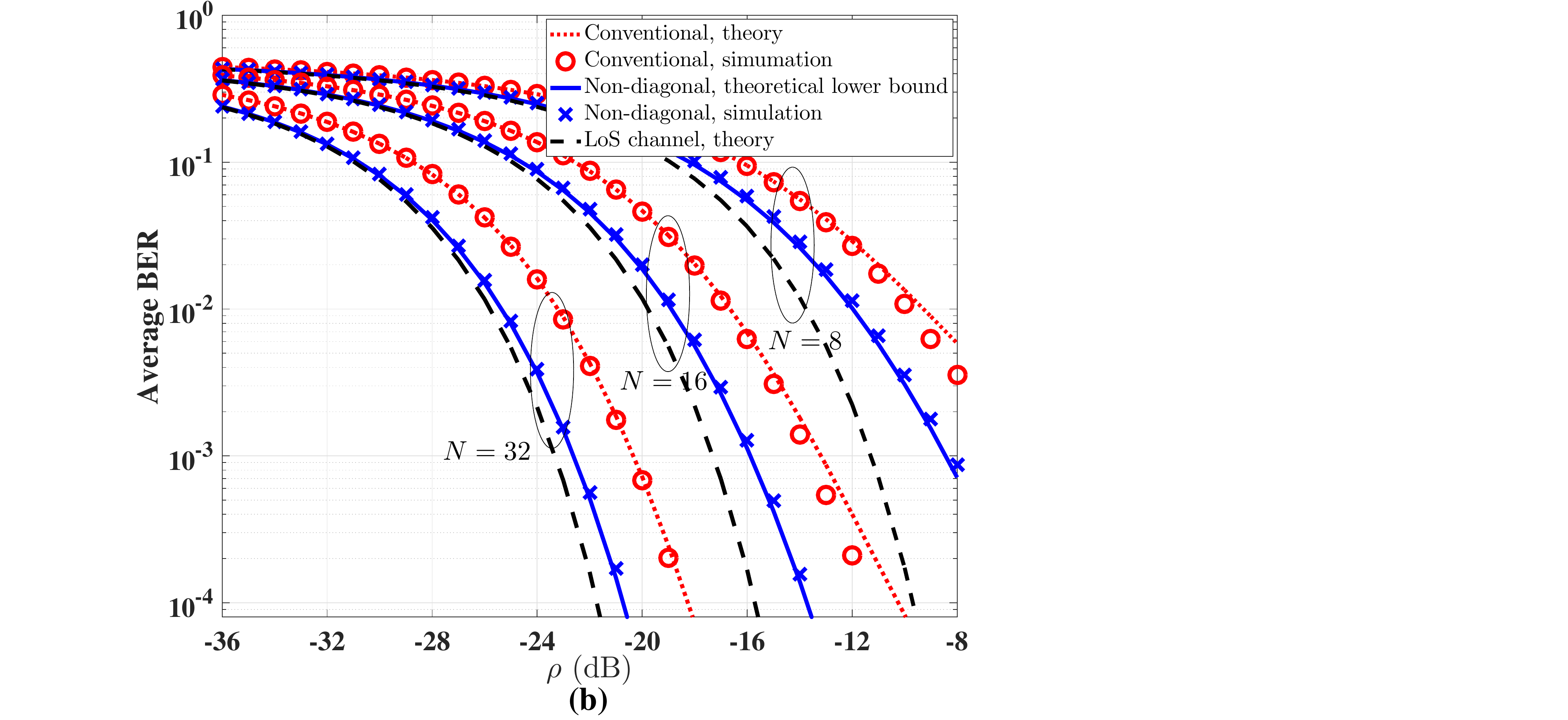}
    \end{minipage}
    \caption{(a) Theoretical lower bound and simulation results of outage probability versus the number of RIS elements $N$ in conventional RIS structure and our proposed RIS structure. (b) Theoretical analysis and simulation results of average BER versus $\rho$ of conventional RIS architecture, our proposed RIS architecture and LoS channel.}\label{Figure_8}
\vspace{-3mm}
\end{figure*}

In Fig. \ref{Figure_8} (a), we show the theoretical lower bound of the outage probability of our proposed RIS architecture in (\ref{Receiver_SNR_6}) versus the number of RIS elements, where the SNR threshold $\omega_{\mathrm{th}}$ is set to $25\mathrm{dB}$. Furthermore, the simulation results of our proposed RIS architecture and of the conventional RIS architecture are presented. Note that the theoretical lower bound is very tight, and the performance of our proposed RIS architecture is significantly better than that of the conventional RIS architecture for all values of $N$. Then, in Fig. \ref{Figure_8} (b) we show that our simulation results and the theoretical average BER of our proposed RIS architecture based on (\ref{Receiver_SNR_7}) match well. The results are contrasted to those using the conventional RIS architecture and to those of the LoS channel, where BPSK modulation is employed. Observe that the average BER performance of our proposed RIS architecture is better than that of the conventional RIS architecture, and tends to that of the LoS channels upon increasing the number of RIS elements.

\begin{figure*}[htb]
\vspace{-0mm}
\setstretch{0.8}
\setlength{\abovecaptionskip}{-3pt}
\setlength{\belowcaptionskip}{-3pt}
    \begin{minipage}[t]{0.33\linewidth}
        \centering
        \includegraphics[width=1.78\textwidth]{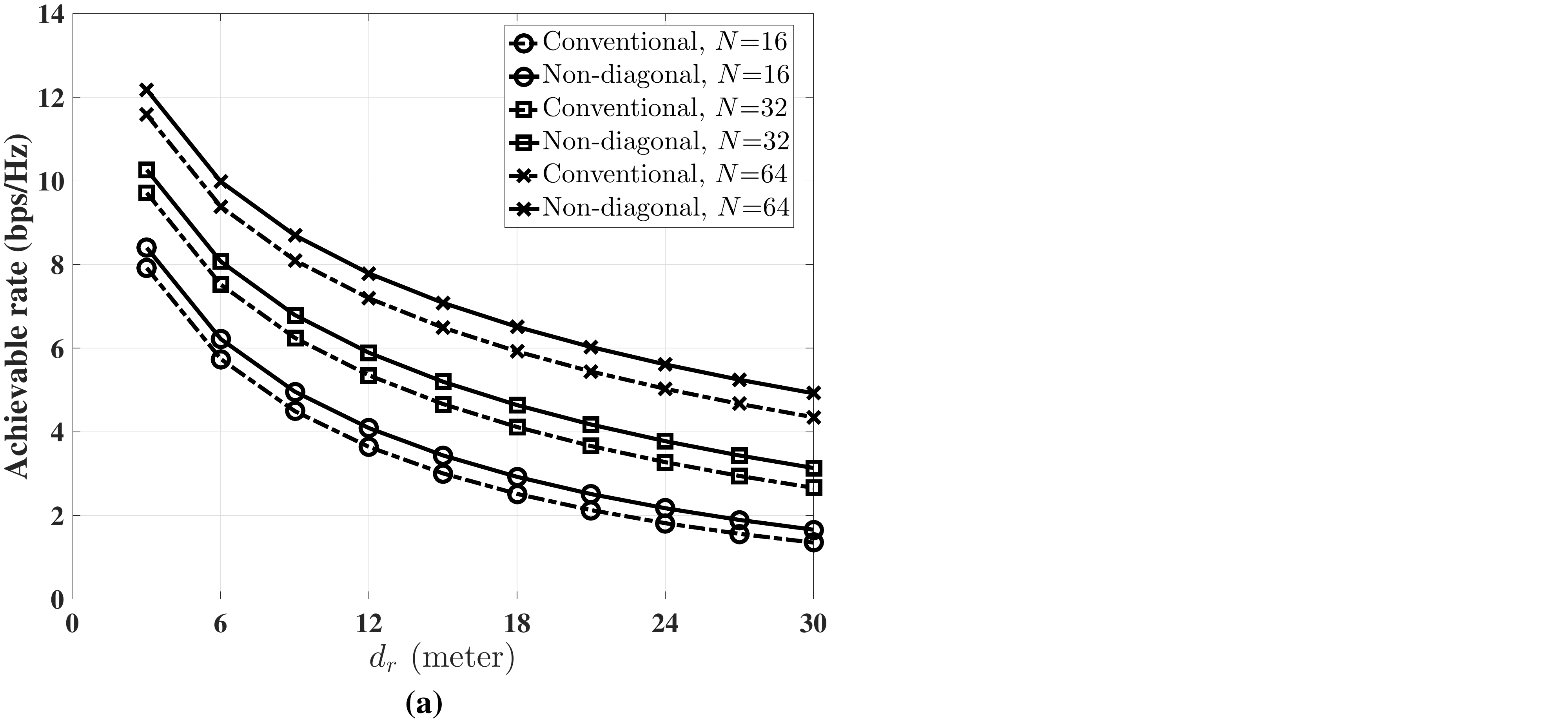}
    \end{minipage}
    \begin{minipage}[t]{0.33\linewidth}
        \centering
        \includegraphics[width=1.78\textwidth]{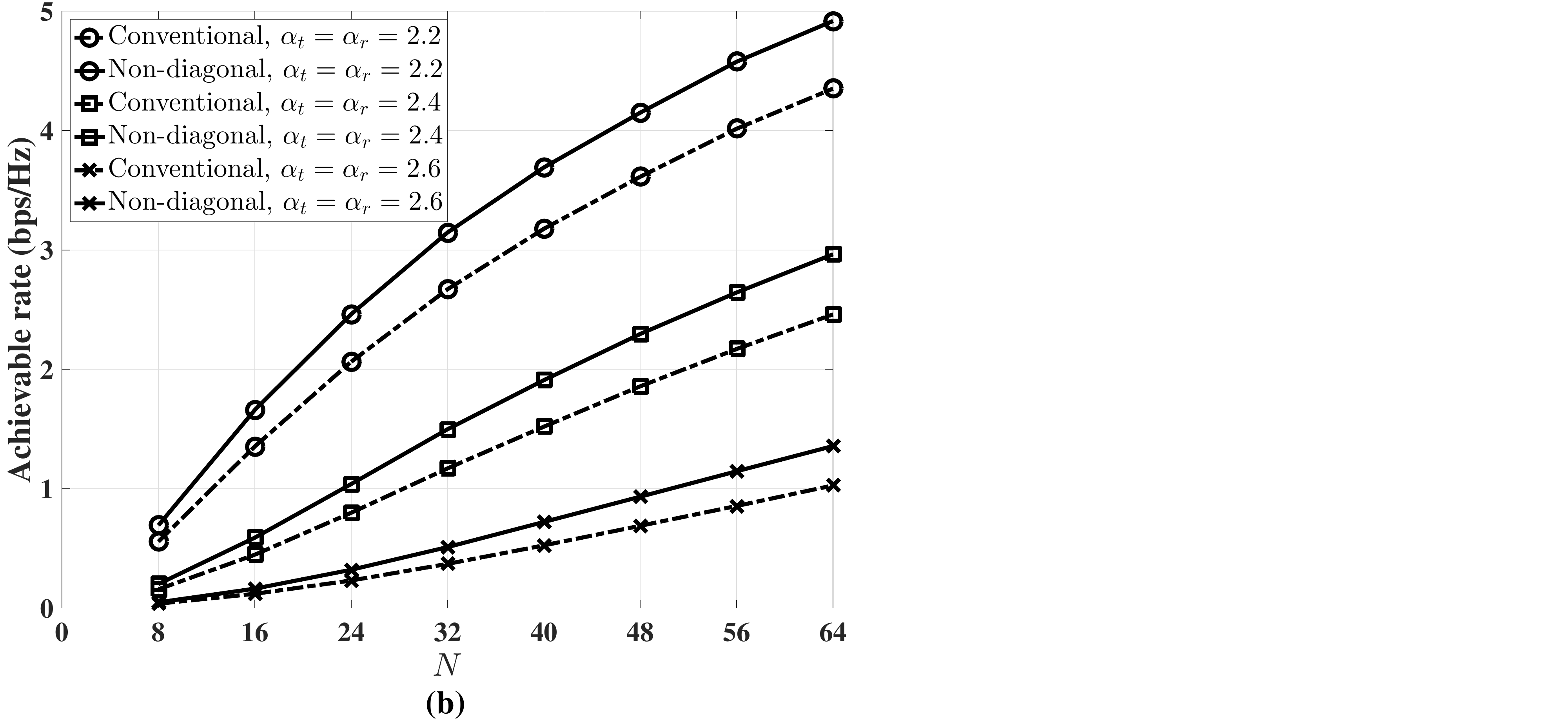}
    \end{minipage}
    \begin{minipage}[t]{0.33\linewidth}
        \centering
        \includegraphics[width=1.78\textwidth]{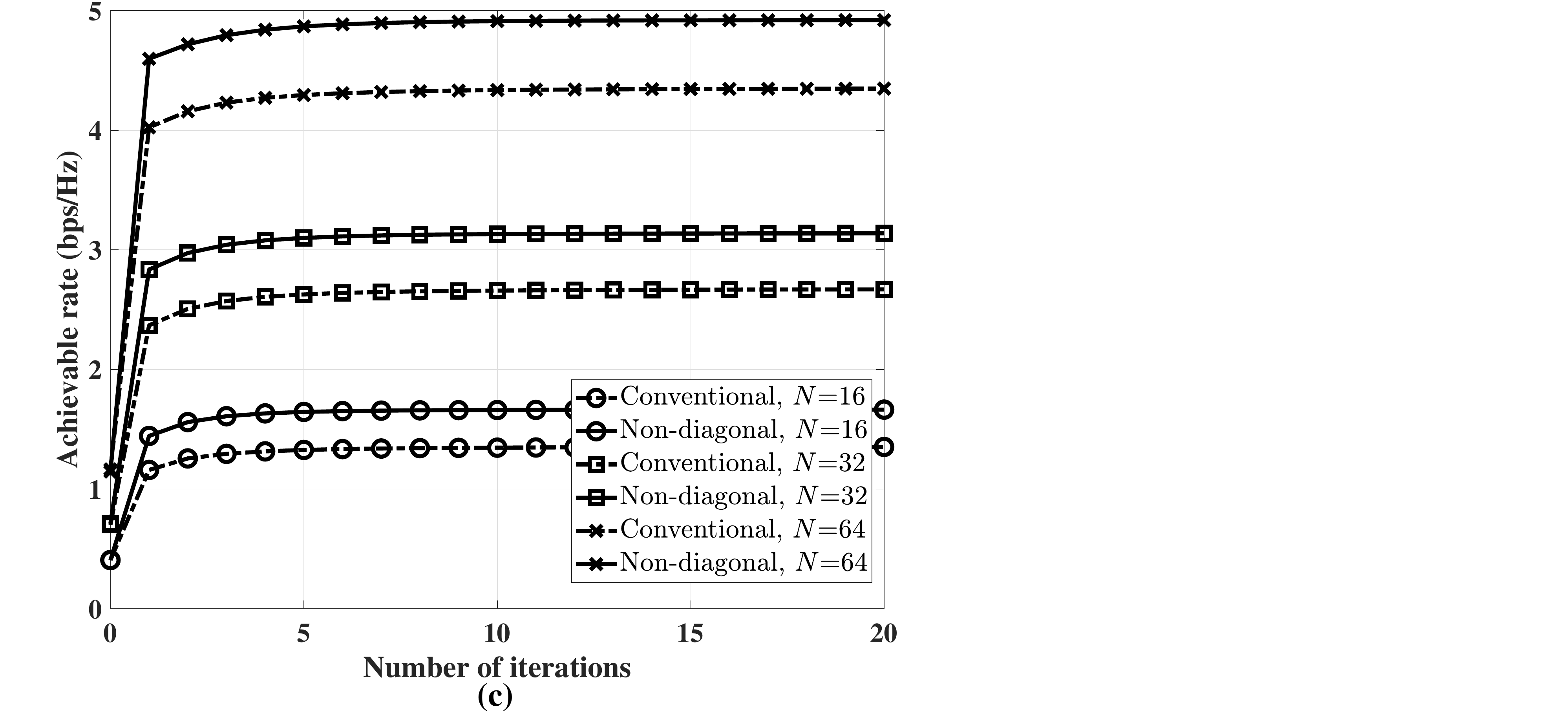}
    \end{minipage}
    \caption{Comparison of the achievable rate in conventional RIS architecture and our proposed RIS architecture in single-user MISO systems: (a) achievable rate versus RIS-user distance $d_r$, (b) achievable rate versus the number of RIS elements $N$, and (c) achievable rate versus the number of iterations in alternating optimization method.}\label{Figure_9}
\vspace{-5mm}
\end{figure*}

\begin{figure*}[htb]
\vspace{-0mm}
\setstretch{0.8}
\setlength{\abovecaptionskip}{-3pt}
\setlength{\belowcaptionskip}{-3pt}
    \begin{minipage}[t]{0.33\linewidth}
        \centering
        \includegraphics[width=1.772\textwidth]{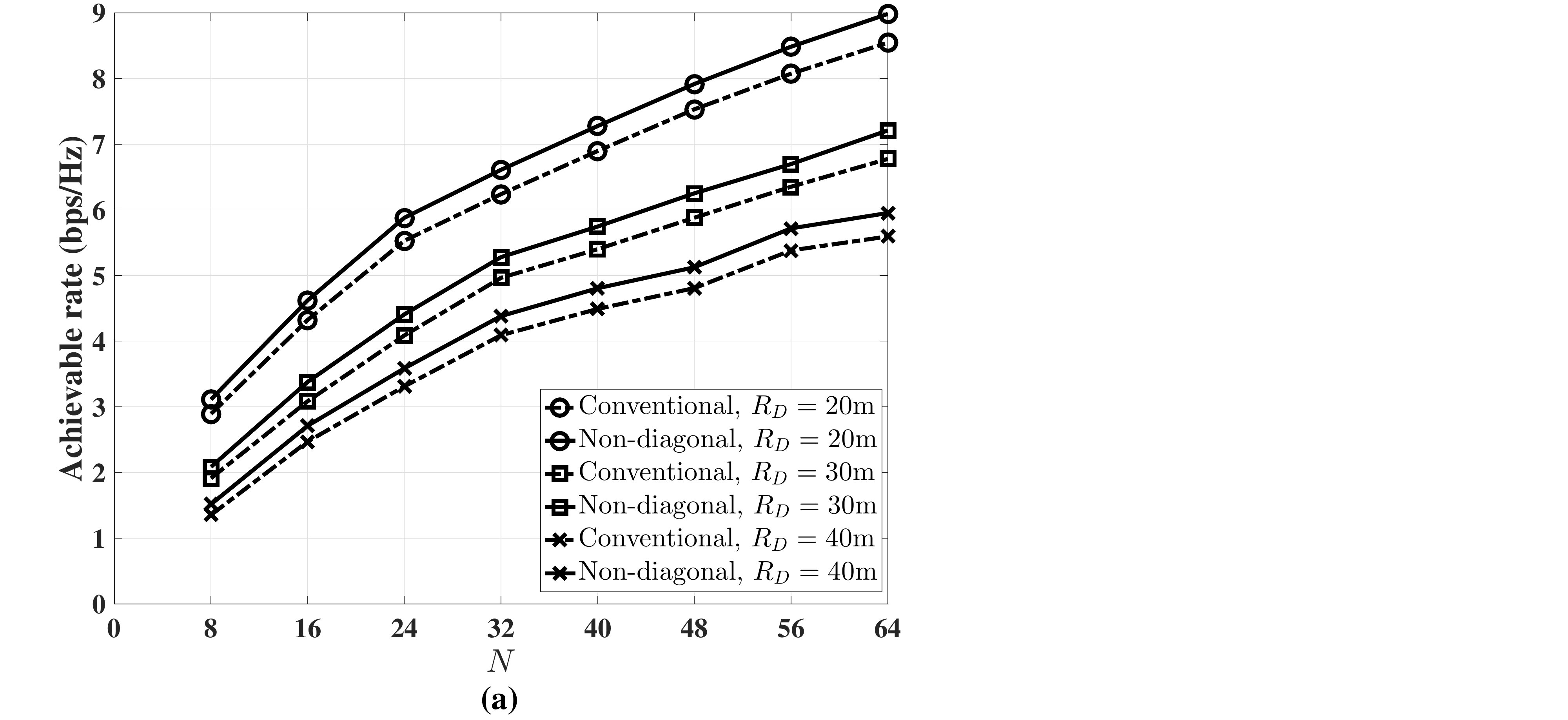}
    \end{minipage}
    \begin{minipage}[t]{0.33\linewidth}
        \centering
        \includegraphics[width=1.78\textwidth]{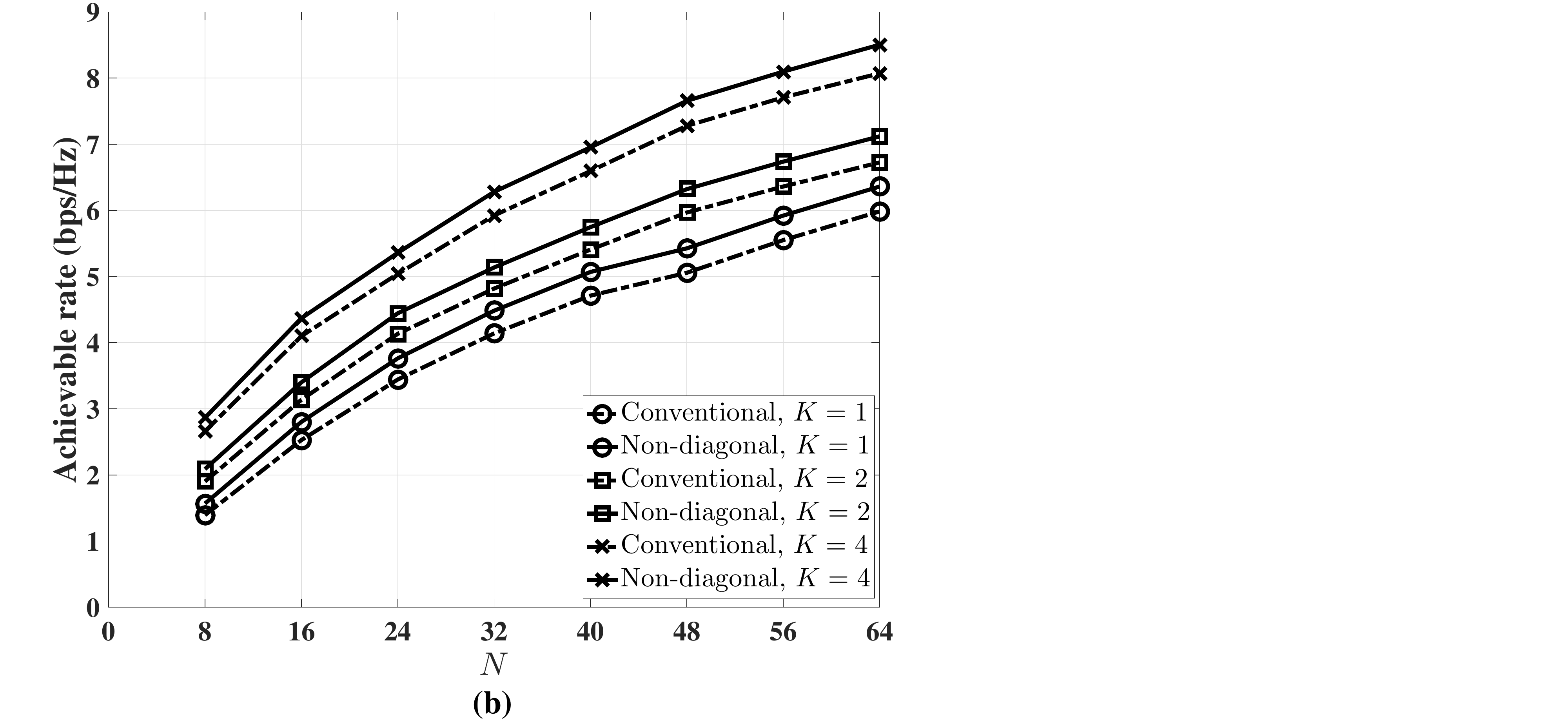}
    \end{minipage}
    \begin{minipage}[t]{0.33\linewidth}
        \centering
        \includegraphics[width=1.78\textwidth]{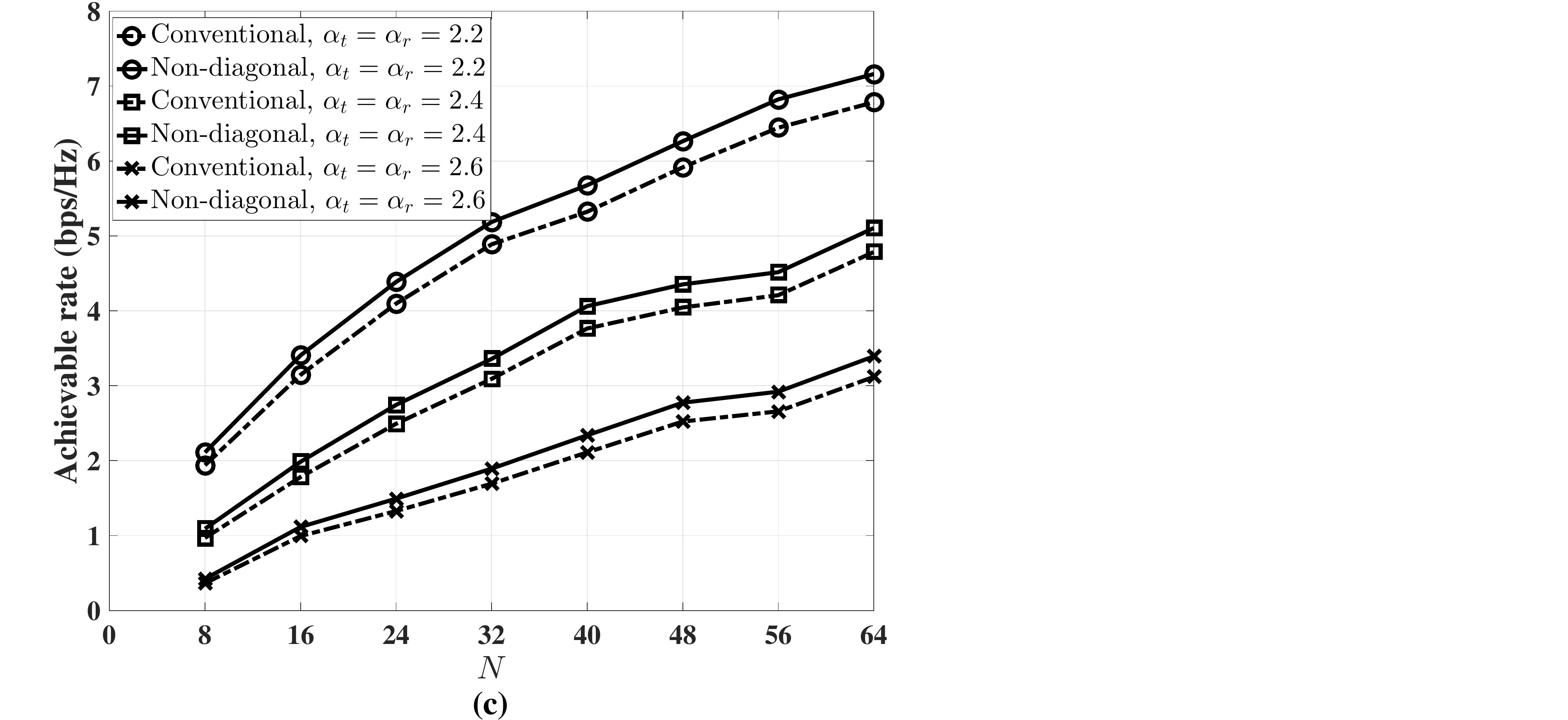}
    \end{minipage}
    \begin{minipage}[t]{0.33\linewidth}
        \centering
        \includegraphics[width=1.78\textwidth]{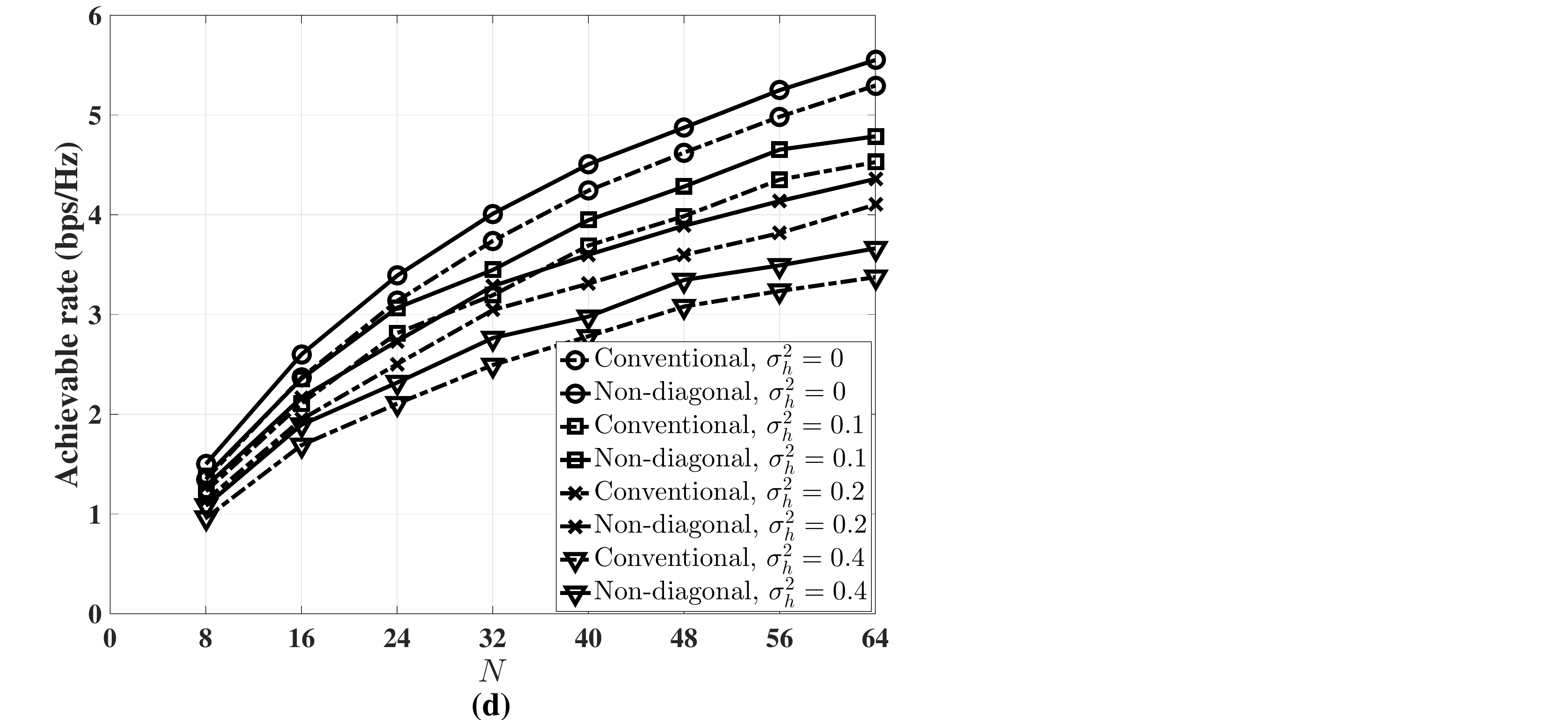}
    \end{minipage}
    \begin{minipage}[t]{0.33\linewidth}
        \centering
        \includegraphics[width=1.78\textwidth]{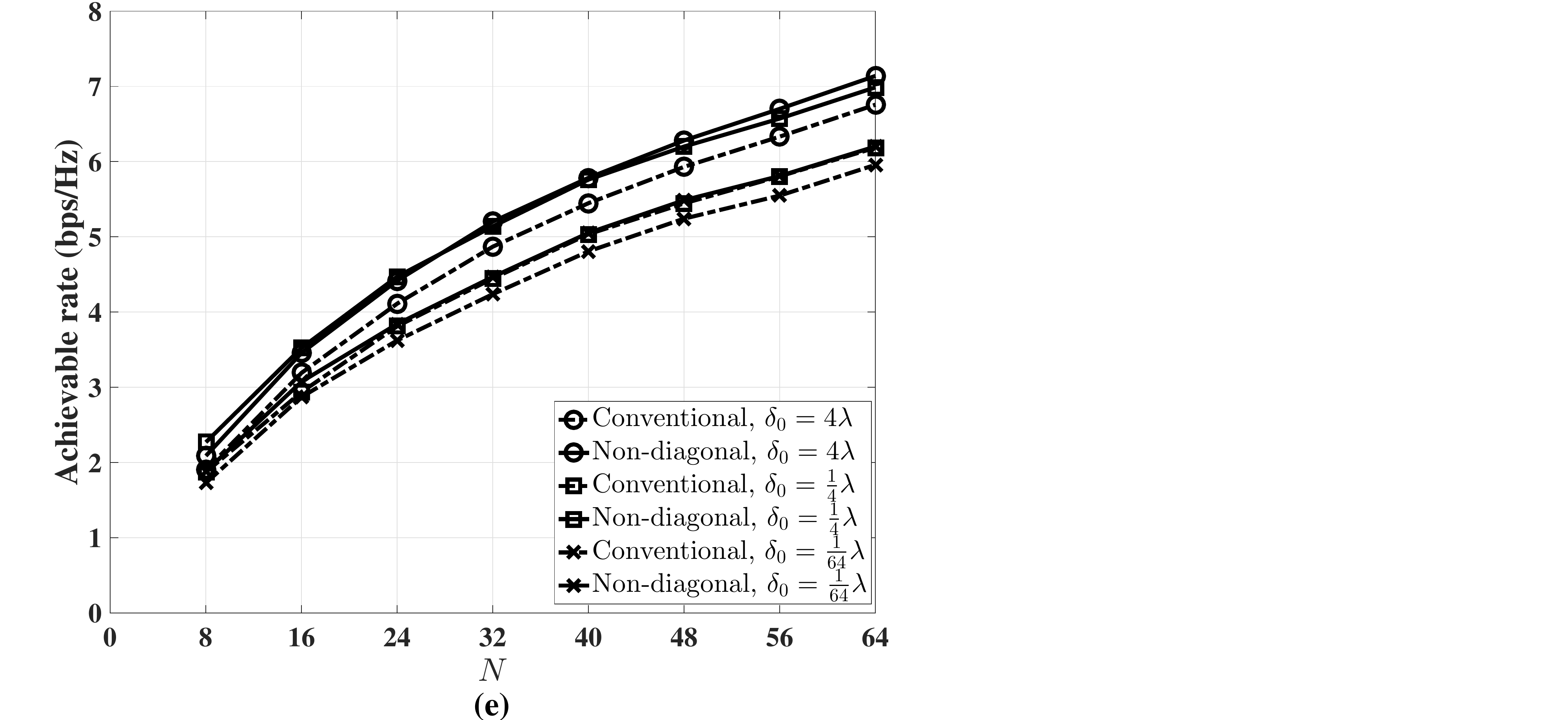}
    \end{minipage}
    \begin{minipage}[t]{0.33\linewidth}
        \centering
        \includegraphics[width=1.78\textwidth]{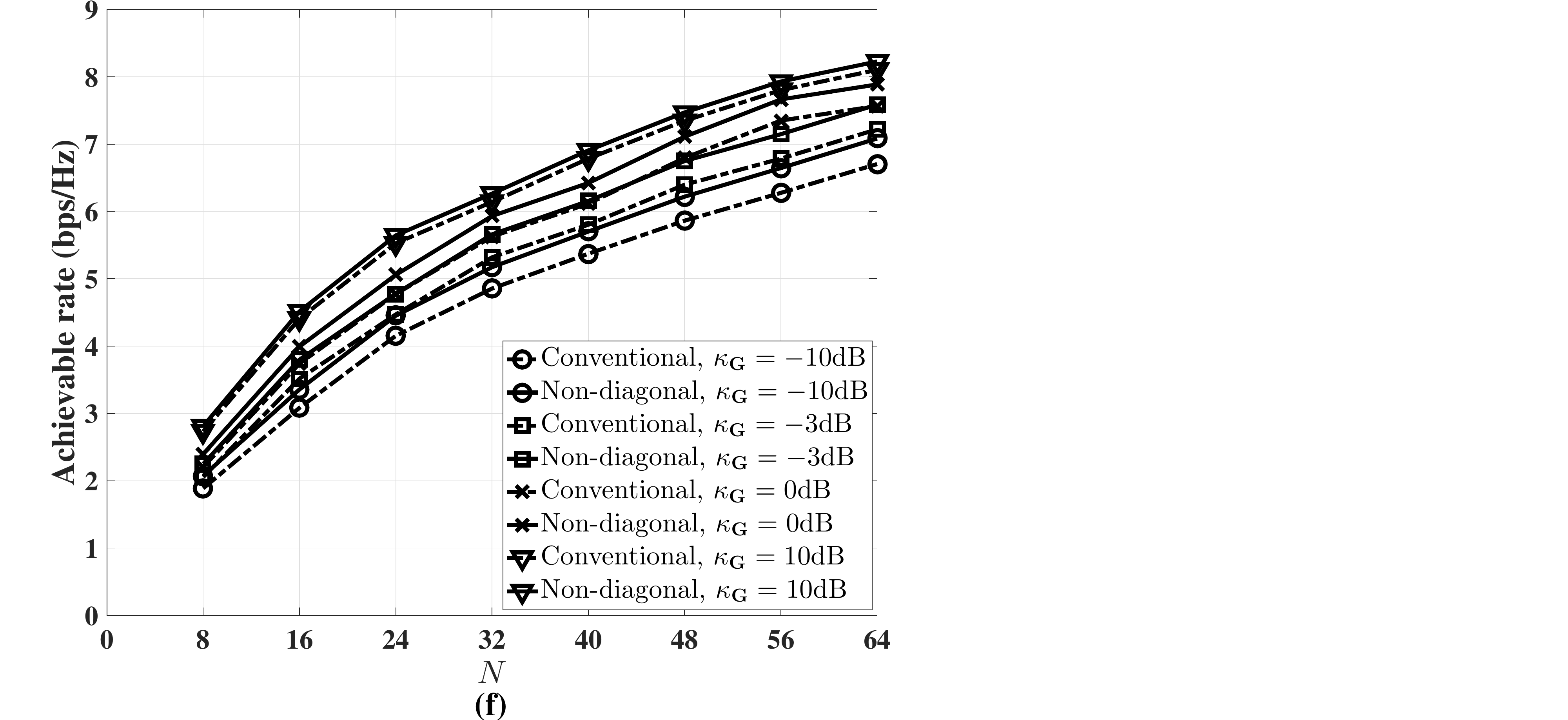}
    \end{minipage}
    \caption{Comparison of the achievable rate versus the number of RIS elements $N$ in conventional RIS structure and our proposed RIS structure in multi-user MIMO systems: (a) with the user radius $R_D=20\mathrm{m},30\mathrm{m},40\mathrm{m}$, (b) with the number of users $K=1,2,4$, (c) with the path loss exponent $\alpha=2.2,2.4,2.6$, (d) with channel estimation error variance $\sigma_h^2=0,0.1,0.2,0.4$, (e) with the distance between adjacent RIS elements $\delta_0=4\lambda,\frac{1}{4}\lambda,\frac{1}{64}\lambda$, and (f) with Rician factor $\alpha_\mathbf{G}=-10\mathrm{dB},-3\mathrm{dB},0\mathrm{dB},10\mathrm{dB}$.}\label{Figure_10}
\vspace{-5mm}
\end{figure*}

In the following simulation, the number of antennas at the BS is $M=4$, the BS-RIS distance $d_t$ is fixed as $d_t=50\mathrm{m}$, the path loss at the reference distance 1 meter is $C_0=-30\mathrm{dB}$, the total transmit power is $P_t=50\mathrm{mW}$, and the noise power is $\sigma_n^2=-90\mathrm{dBm}$. The Rician factors are $\kappa_{\mathbf{G}}=\kappa_{\mathbf{h}_1}=\kappa_{\mathbf{h}_2}=\cdots=\kappa_{\mathbf{h}_K}=-10\mathrm{dB}$, unless otherwise specified.

Firstly, the achievable rate of single-user MISO systems is presented based on (\ref{beamforming_design_miso_3_1}) in Fig. \ref{Figure_9} (a), where we compare the achievable rate versus RIS-user distance $d_r$ of both the conventional RIS and of our proposed RIS structure. The path loss exponent is $\alpha_t=\alpha_r=2.2$. Observe that the achievable rate decreases upon the increasing the RIS-user distance due to the increased path loss. We can also find that the achievable rate of our proposed RIS architecture is higher than that of the conventional RIS architecture for all RIS-user distances. In Fig. \ref{Figure_9} (b), we compare the achievable rate versus the number of RIS elements $N$ in the conventional RIS architecture and our proposed RIS architecture, where the RIS-user distance is $d_r=30\mathrm{m}$. Observe that the performance of our proposed RIS architecture is better than that of the conventional RIS architecture, especially when the number of RIS elements $N$ is large. This is due to the fact that our proposed RIS architecture benefits from using the MRC criterion upon increasing the number of RIS elements. In Fig. \ref{Figure_9} (c), the achievable rate versus the number of iterations in alternating optimization algorithm is presented, where the path loss exponent $\alpha_t=\alpha_r=2.2$ and RIS-user distance is $d_r=30\mathrm{m}$. It shows that the convergence speed in our proposed RIS architecture is almost the same as that of the conventional RIS architecture.

Then, the simulation based on achievable rate of the multi-user MIMO systems is presented in Fig. \ref{Figure_10}, where the users are distributed uniformly within a half circle centered at the RIS with radius $R_D$. Explicitly, we compare the achievable rate versus the number of RIS elements $N$ of the conventional and of our proposed RIS architecture. In Fig. \ref{Figure_10} (a), the path loss exponent is $\alpha_t=\alpha_r=2.2$, and the number of users is $K=2$. It can be seen that compared to the conventional RIS architecture, our proposed RIS architecture has better coverage under the same data rate requirement. In Fig. \ref{Figure_10} (b), the path loss exponent is $\alpha_t=\alpha_r=2.2$, and the radius is $R_D=30\mathrm{m}$, where it can be seen that the achievable rate increases near linearly with the number of RIS elements. Compared to the conventional RIS architecture, the achievable rate of our proposed method is significantly high for any number of users. In Fig. \ref{Figure_10} (c), the number of users is $K=2$, and the radius is $R_D=30\mathrm{m}$. We can observe the performance enhancement of our proposed RIS architecture when the number of RIS elements $N$ is large for any pathloss exponent value.

When considering imperfect CSI, referring to \cite{yoo2004capacity}, the estimated BS-RIS and RIS-user channels are given by $\hat{\mathbf{g}}_m=\mathbf{g}_m+\mathbf{g}_{m,e}$ and $\hat{\mathbf{h}}_k^{\mathrm{H}}=\mathbf{h}_k^{\mathrm{H}}+\mathbf{h}_{k,e}^{\mathrm{H}}$, respectively, where $m=1,2,\cdots,M$, $k=1,2,\cdots,K$. The channel estimation error component obeys $\mathbf{g}_{m,e}\sim\mathcal{CN}\left(\mathbf{0},\sigma_h^2\mathbf{I}\right)$, $\mathbf{h}_{k,e}^{\mathrm{H}}\sim\mathcal{CN}\left(\mathbf{0},\sigma_h^2\mathbf{I}\right)$, where $\sigma_h^2$ represents the estimation error variance. In the linear minimum mean squared error (LMMSE) method, we have $\sigma_h^2=\frac{1}{1+\tau_p\rho_t}$, where $\rho_t$ is the pilot transmission power and $\tau_p$ represents the pilot symbols¡¯ transmission duration \cite{katla2020deep}. In Fig. \ref{Figure_10} (d), the achievable rate is presented when considering imperfect CSI, where the path loss exponent is $\alpha_t=\alpha_r=2.2$, the number of users is $K=2$ and the radius is $R_D=30\mathrm{m}$. As shown in Fig. \ref{Figure_10} (d), our proposed RIS architecture outperforms the conventional RIS architecture under any channel estimation error variance $\sigma_h^2$, which shows the robustness of our proposed architecture compared to the conventional one.

In most of the previous contributions \cite{wu2019intelligent,ning2020beamforming,wang2020intelligent,han2019large,wu2019beamforming,zhang2020beyond,chen2019sum,xu2019discrete,lin2020reconfigurable,basar2019wireless,yang2020accurate,shen2020modeling}, it is assumed that the distance between the adjacent RIS elements is large enough to ensure that the BS-RIS and RIS-user channels are uncorrelated. To make our work more realistic, in Fig. \ref{Figure_10} (e), we employ the exponential correlation channel model of \cite{hampton2013introduction} to investigate the effect of channel correlation on the achievable rate, where the path loss exponent is $\alpha_t=\alpha_r=2.2$, the number of users is $K=2$, the radius is $R_D=30\mathrm{m}$ and the reference correlation distance $d_{\mathrm{ref}}=\lambda$. As shown in Fig. \ref{Figure_10} (e), our proposed RIS architecture attains a better performance than that of the conventional RIS architecture. For example, the achievable rate in our proposed RIS architecture with adjacent RIS element distance $\delta_0=\frac{1}{64}\lambda$ is almost the same as that in conventional RIS architecture with adjacent RIS element distance $\delta_0=\frac{1}{4}\lambda$, and the achievable rate in our proposed RIS architecture with adjacent RIS element distance $\delta_0=\frac{1}{4}\lambda$ is even higher than that in conventional RIS architecture with adjacent RIS element distance $\delta_0=4\lambda$.

Finally, we investigate the impact of the Rician factor on the achievable rate in Fig. \ref{Figure_10} (f), where the path loss exponent is $\alpha_t=\alpha_r=2.2$, the number of users is $K=2$, the radius is $R_D=30\mathrm{m}$. We also assume the Rician factor of RIS-user path is $\kappa_{\mathbf{h}_r}=-20\mathrm{dB}$ for $k=1,2,\cdots,K$, while the Rician factor of BS-RIS path $\kappa_{\mathbf{G}}$ can range between $-10\mathrm{dB}$ to $10\mathrm{dB}$. With the increase of the Rician factor $\kappa_{\mathbf{G}}$, the achievable rate of the conventional RIS architecture and that of our proposed RIS architecture can both achieve better performance, which is due to the reduced impact of channel fading with the increase of LoS component. Besides, it shows that the achievable rate enhancement of our proposed RIS architecture is not obvious in the high range of Rician factor, while it is significant when the channel is dominated by NLoS component, which agrees with the above theoretical analysis.

\section{Conclusions}\label{Section_Conclusions}
In this paper, we proposed a novel RIS architecture, where the signal impinging on a specific element can be reflected by another element after phase shift adjustment. Compared to the conventional RIS architecture, our proposed RIS architecture provides better channel gain as a benefit of using the MRC criterion. The theoretical analysis showed that our system performs better than the conventional RIS system both in terms of its average channel gain, outage probability and average BER. Furthermore, the performance of our proposed RIS architecture is better than that of the group-connected RIS architecture and approaches that of the fully-connected RIS architecture, with increasing the number of RIS elements, despite its considerably reduced complexity compared to the fully-connected architecture. Furthermore, we formulated and solved the problem of maximizing the achievable rate of our proposed RIS architecture by jointly optimizing the active TBF and the non-diagonal phase shift matrix of both single-user MISO systems and of multi-user MIMO systems based on alternating optimization and on SDR methods, respectively. The simulation results showed that our proposed technique is capacity of enhancing the achievable rate of RIS-assisted wireless communications.

\appendices

\section{Proof of formula (\ref{K_Effect_0})}\label{Appendix_1}
When $\kappa_{\mathbf{g}}\rightarrow\infty$, according to (\ref{channel_gain_1}), we can get that $a_i=1$ ($i=1,2,\cdots,N$). Therefore,
\begin{align}\label{K_Effect_1}
    \sum_{i=1}^{N}a_ib_i=\sum_{i=1}^{N}a_{(i)}b_{(i)}=\sum_{i=1}^{N}b_i.
\end{align}
According to (\ref{Rician_scaling_law_conven_1}), (\ref{Rician_scaling_law_sort_1}) and ({\ref{K_Effect_1}}), we can show that $\mathfrak{g}_\mathrm{diag}=\mathfrak{g}_\mathrm{nond}$. When we consider the case that $\kappa_{\mathbf{h}}\rightarrow\infty$, we can similarly get that $\mathfrak{g}_\mathrm{diag}=\mathfrak{g}_\mathrm{nond}$. The proof is thus completed.

\section{Proof of \textit{Theorem} \ref{theorem_1}}\label{Appendix_2}
We denote the CDF of the random variable $X$ following the Rayleigh distribution with the scale parameter $\sigma$ as $F_{\mathrm{X}}(x;\sigma)$. The PDF of $a_{(i)}$ is given by \cite{renyi1953theory}
\begin{align}\label{scaling_law_proposed_5}
    \notag f_{a_{(i)}}(x)&=C_{i,N}f_{\mathrm{X}}\left(x;\sigma=\frac{\sqrt{2}}{2}\right)\left[F_{\mathrm{X}}\left(x;\sigma=\frac{\sqrt{2}}{2}\right)\right]^{i-1}\times\\
    \notag&\quad\left[1-F_{\mathrm{X}}\left(x;\sigma=\frac{\sqrt{2}}{2}\right)\right]^{N-i}\\
    \notag&=2C_{i,N}xe^{-x^2}\left(1-e^{-x^2}\right)^{i-1} \left(e^{-x^2}\right)^{N-i}\\
    &=2C_{i,N}x\left(1-e^{-x^2}\right)^{i-1}\left(e^{-x^2}\right)^{N-(i-1)}.
\end{align}
In (\ref{scaling_law_proposed_5}), $\left(1-e^{-x^2}\right)^{i-1}=\sum_{k=0}^{i-1}\binom{i-1}{k}\left(-e^{-x^2}\right)^k$, so
\begin{align}\label{scaling_law_proposed_6}
    \notag  f_{a_{(i)}}(x)&=2C_{i,N}x\left(e^{-x^2}\right)^{N-(i-1)}\sum_{k=0}^{i-1}\binom{i-1}{k}\left(-e^{-x^2}\right)^k\\
    \notag&=2C_{i,N}\sum_{k=0}^{i-1}\binom{i-1}{k}x\left(-e^{-x^2}\right)^{N-(i-1)+k}\\
    \notag&=C_{i,N}\sum_{k=0}^{i-1}\binom{i-1}{k}\frac{1}{N-i+k+1}(-1)^{k}\times\\
    \notag&\qquad f_{\mathrm{X}}\left(x;\sigma=\frac{1}{\sqrt{2(N-i+k+1)}}\right)\\
    \notag&=\sum_{k=0}^{i-1}\binom{N}{i-k-1}\binom{N-i+k}{k}(-1)^k\times\\
    &\qquad f_{\mathrm{X}}\left(x;\sigma=\frac{1}{\sqrt{2(N-i+k+1)}}\right).
\end{align}
The proof is thus completed.

\section{Proof of formula (\ref{N_infty_2})}\label{Appendix_3}
In (\ref{Rician_scaling_law_sort_1}), according to Cauchy-Schwarz inequality, we can get
\begin{align}\label{N_infty_3}
    \notag \mathfrak{g}_\mathrm{nond}&= \varrho_t\varrho_r \mathbb{E}\left(\left(\sum_{i=1}^{N}a_{(i)}b_{(i)}\right)^2\right)\\
    &\leq\varrho_t\varrho_r\mathbb{E}\left(\left(\sum_{i=1}^{N}a_{(i)}^2\right)\left(\sum_{i=1}^{N}b_{(i)}^2\right)\right),
\end{align}
where the equality in (\ref{N_infty_3}) is established only when $a_{(i)}=b_{(i)}$. When $N\rightarrow\infty$, we can get
\begin{align}\label{N_infty_4}
    a_{(i)}=b_{(i)}=F_{a_{i}}^{-1}\left(\frac{i}{N}\right)=F_{b_{i}}^{-1}\left(\frac{i}{N}\right),
\end{align}
where $F_{a_{i}}^{-1}(\cdot)$ and $F_{b_{i}}^{-1}(\cdot)$ represent the inverse CDF of $a_{(i)}$ and $b_{(i)}$, respectively. Based on (\ref{N_infty_3}) and (\ref{N_infty_4}), we can show that when $N\rightarrow\infty$, the channel gain of our proposed RIS architecture is
\begin{align}\label{N_infty_5}
    \notag \mathfrak{g}_\mathrm{nond}&=\varrho_t\varrho_r\mathbb{E}\left(\left(\sum_{i=1}^{N}a_{(i)}^2\right)\left(\sum_{i=1}^{N}b_{(i)}^2\right)\right)\\
    \notag&=\varrho_t\varrho_r\mathbb{E}\left(\sum_{i=1}^{N}a_i^2\right)\mathbb{E}\left(\sum_{i=1}^{N}b_i^2\right)\\
    &=\varrho_t\varrho_rN^2.
\end{align}
The proof is thus completed.

\bibliographystyle{IEEEtran}
\bibliography{IEEEabrv,TAMS}
\end{document}